\documentclass[12pt]{article}
\usepackage[cp1250]{inputenc}
\usepackage[verbose,a4paper,tmargin=0.5in,lmargin=1in,rmargin=1in]{geometry}
\usepackage[pdftex]{graphicx}
\usepackage{amsfonts}
\usepackage{indentfirst}
\usepackage{latexsym}
\usepackage{amsmath}
\usepackage{amssymb}
\usepackage{psfrag}
\usepackage{eufrak}
\usepackage[mathscr]{eucal} 

\renewcommand{\theequation}{\arabic{section}.\arabic{equation}}

\title{Killing Symmetries in $\mathcal{H}$-Spaces with $\Lambda$.}

\author{$\textrm{Adam Chudecki}^{*} \dag$
\ \ and \ \  $\textrm{Maciej Przanowski}^{**} \ddag$}

\begin{document}

\maketitle

$*$ Center of Mathematics and Physics  

$\ \ $ Technical University of Łódź           

$\ \ $ Al. Politechniki 11, 90-924 Łódź, Poland
\newline

$**$ Institute of Physics  

$\ \ $ Technical University of Łódź           

$\ \ $ Wólczańska 219, 90-924 Łódź, Poland
\newline

$\dag$ E-mail addres: adam.chudecki@p.lodz.pl

$\ddag$ E-mail addres: maciej.przanowski@p.lodz.pl
\newline
\newline
\textbf{Abstract}. All Killing symmetries in complex $\mathcal{H}$-spaces with $\Lambda$ in terms of the Plebański - Robinson - Finley coordinate system are found. All $\mathcal{H}$-metrics with $\Lambda$ admitting a null Killing vector are explicitly given. It is shown that the problem of non-null Killing vector reduces to looking for solution of the \textsl{Boyer - Finley - Plebański (Toda field)} equation.

\section{Introduction}

When in 1976 the famous paper by J.F. Plebański and I. Robinson on the hyperheavenly spaces appeared \cite{biblio_1} it was quite clear for all who were interested in heavenly spaces and complex methods in relativity that a new powerful tool has been just introduced. Further developement of this subject by J.F. Plebański, I. Robinson, C.P. Boyer, J.D. Finley III, T. del Castillo, A. Garcia, K. Rózga and many, many others provided us with deep understanding of a broad application of complex analysis in relativity \cite{biblio_2} - \cite{biblio_dod_2}. Later on it has been realized that the hyperheavenly spaces play also a crucial role in the theory of Osserman and Walker spaces \cite{biblio_9, biblio_10, biblio_dod_1}.

One of the problems which, from the very beginning, has attracted a great deal of interest was the problem of classifying all $\mathcal{H}$ and $\mathcal{HH}$-spaces admitting (conformal, homothetic or isometric) Killing vector \cite{biblio_7} - \cite{biblio_dod_2}, \cite{biblio_15} - \cite{biblio_17}. It seems that the only case which has been not analysed in that context is the case of complex $\mathcal{H}$-space with $\Lambda$.

The aim of the present paper is to fill this gap and to classify all $\mathcal{H}$-spaces with $\Lambda$ admitting a Killing vector. We intend to give this classification using the coordinate system explored intensively in the $\mathcal{HH}$-space theory \cite{biblio_1} - \cite{biblio_4}. Although, as has been shown by one of us \cite{biblio_11}, P. Tod \cite{biblio_12} and M. H\"ogner \cite{biblio_13} in the case of real $\mathcal{H}$-spaces with $\Lambda$ admitting a Killing vector some other coordinate systems are more convenient, we intend to show in the present work that in the complex case the Plebański - Robinson - Finley coordinates are still extraordinary useful and transparent.

Our paper is organized as follows. 

Section 2 is devoted to general properties of $\mathcal{H}$-spaces and $\mathcal{HH}$-spaces with $\Lambda$. Heavenly equation with $\Lambda$ is given and the gauge freedom is studied. In section 3 the Killing equations and their integrability conditions are considered. The general form of the Killing vector is found (see (\ref{ostateczna_postac_wektora_Kili})) and the master equation (\ref{expanding_master_equation}) is derived. Transformation rules of the components of the Killing vector are given.

The main results of the paper are gathered in section 4 where the classification of all $\mathcal{H}$-spaces with $\Lambda$ admitting Killing symmetries are found. In subsection 4.1 some important lemmas are proved which are then extensively used to find the general metric (\ref{main_metryka_null}) of $\mathcal{H}$-space with $\Lambda$ admitting a null Killing vector.

The general form of non-null Killing vector (\ref{nonnnnulll_Kil}) is obtained in subsection 4.3 and it is shown that the problem is reduced to finding the solution of the Boyer - Finley - Plebański (Toda field) equation (\ref{rownanie_BFP}).

Details concerning the calculations can be found in Appendix A.

\section{Heavenly spaces ($\mathcal{H}$-spaces) with $\Lambda$.}

\subsection{General structure of hyperheavenly spaces ($\mathcal{HH}$-spaces) with $\Lambda$.}
\label{etykietka_subsekcji_21}

\textsl{$\mathcal{HH}$-space with cosmological constant $\Lambda$} is a 4 - dimensional complex analytic differential manifold endowed with a holomorphic Riemannian metric $ds^{2}$ satisfying the vacuum Einstein equations with $\Lambda$, $R_{ab}=-\Lambda g_{ab}$, $\Lambda \ne 0$, and such that the self - dual or anti - self - dual part of the Weyl tensor is algebraically degenerate \cite{biblio_1} - \cite{biblio_4}. This kind of spaces admits a congruence of totally null, self-dual (or anti-self-dual, respectively) surfaces, called \textsl{null strings} or \textsl{twistor surfaces} \cite{biblio_14} - \cite{biblio_dod_5}. In what follows we deal with the case when the self-dual part of Weyl tensor is degenerate. Then, a null tetrad $(e^{1},e^{2},e^{3},e^{4})$ and a coordinate system $(q_{\dot{A}}, p^{\dot{B}})$ can be always chosen so that $\Gamma_{422}=\Gamma_{424}=0$ \cite{biblio_2}, the surface element of the self-dual null string is given by $e^{1} \wedge e^{3}$, and the null tetrad $(e^{1},e^{2},e^{3},e^{4})$ in spinorial notation reads
\begin{eqnarray}
\label{tetrada_spinorowa}
e_{\dot{A}} := \phi^{-2} \, dq_{\dot{A}} &=& 
\left[ \begin{array}{c}  e^{3}  \\ e^{1}  \end{array} \right]  \ \ = \ \ 
-\frac{1}{\sqrt{2}} \, g^{2}_{\ \, \dot{A}}
\\ \nonumber
E^{\dot{A}} := -dp^{\dot{A}} + Q^{\dot{A} \dot{B}} \, dq_{\dot{B}} &=&
\left[ \begin{array}{c}  e^{4}  \\ e^{2}  \end{array} \right]  \ \ = \ \ 
\frac{1}{\sqrt{2}} \, g^{1 \dot{A}} \ \ \ , \ \ \ \ \dot{A}, \dot{B} = \dot{1}, \dot{2}
\end{eqnarray}
where
\begin{equation}
(g^{A\dot{B}}) := \sqrt{2}
\left[\begin{array}{cc}
e^4 & e^2 \\
e^1 & -e^3
\end{array}\right] 
\end{equation}
and $\phi$ and $Q^{\dot{A} \dot{B}}=Q^{\dot{B} \dot{A}}$ are holomorphic functions. Spinorial coordinates $p^{\dot{A}}$ are coordinates on null strings, while $q_{\dot{A}}$ just label them.
Define the following operators
\begin{eqnarray}
&&\partial_{\dot{A}} := \frac{\partial}{\partial p^{\dot{A}}}
\ \ \ \ \ \ \ \ \ \ \ \ \ \ \ \ \ \ \ \ \ \ \ \ 
\partial^{\dot{A}} := \frac{\partial}{\partial p_{\dot{A}}}
\\ \nonumber
&&\eth^{\dot{A}} := \phi^{2} \left( \frac{\partial}{\partial q_{\dot{A}}} + Q^{\dot{A} \dot{B}} \partial_{\dot{B}} \right)
\ \ \ \ \ \ \ \ 
\eth_{\dot{A}} := \phi^{2} \left( \frac{\partial}{\partial q^{\dot{A}}} - Q_{\dot{A}}^{\ \ \dot{B}} \partial_{\dot{B}} \right)
\end{eqnarray}
They constitute the basis dual to $(e_{\dot{A}}, E^{\dot{B}})$
\begin{equation}
\label{baza_dualna}
-\partial_{\dot{A}} = 
\left[ \begin{array}{c}  \partial_{4}  \\ \partial_{2}  \end{array} \right] 
 \ \ \ \ \ \ \ \ 
\eth^{\dot{A}} = 
\left[ \begin{array}{c}  \partial_{3}  \\ \partial_{1}  \end{array} \right] 
\end{equation}
We use the following rules of manipulating the spinorial indices: $q_{\dot{A}} = \ \in_{\dot{A} \dot{B}} q^{\dot{B}}$, $q^{\dot{A}} = q_{\dot{B}} \in^{\dot{B} \dot{A}}$, but for consistency, the rules to raise and lower spinor indices in the case of objects from tangent space read $\partial^{\dot{A}} = \partial_{\dot{B}} \in^{\dot{A} \dot{B}}$, $\partial_{\dot{A}} = \in_{\dot{B} \dot{A}} \partial^{\dot{B}}$, $\eth^{\dot{A}} = \eth_{\dot{B}} \in^{\dot{A} \dot{B}}$, $\eth_{\dot{A}} = \in_{\dot{B} \dot{A}} \eth^{\dot{B}}$, where $\in_{AB}$ and $\in_{\dot{A}\dot{B}}$ are the spinor Levi-Civita symbols
\begin{eqnarray}
&& \in_{AB}  := \left[ \begin{array}{cc}
                            0 & 1   \\
                           -1 & 0  
                            \end{array} \right] =:  \in^{AB} 
\ \ \ , \ \ \ 
 \in_{\dot{A}\dot{B}}  := \left[ \begin{array}{cc}
                            0 & 1   \\
                           -1 & 0  
                            \end{array} \right] =:  \in^{\dot{A}\dot{B}}  
\\ \nonumber
&& \in_{AC} \in^{AB} = \delta^{B}_{C} \ \ \ , \ \ \ \in_{\dot{A}\dot{C}} \in^{\dot{A}\dot{B}} = \delta^{\dot{B}}_{\dot{C}} \ \ \ , \ \ \ 
\delta^{A}_{C}= \delta^{\dot{B}}_{\dot{C}}= \left[ \begin{array}{cc}
                            1 & 0   \\
                            0 & 1  
                            \end{array} \right]
\end{eqnarray}
The metric $ds^2$ is given by
\begin{equation}
\label{metryka_ogolna}
ds^{2} =2 \, e^{1} \underset{s}{\otimes} e^{2} + 2 \, e^{3} \underset{s}{\otimes} e^{4}
= -\frac{1}{2} \, g_{A\dot{B}} \underset{s}{\otimes} g^{A\dot{B}}
= 2\, \phi^{-2} \, (- dp^{\dot{A}} \underset{s}{\otimes} dq_{\dot{A}} + 
           Q^{\dot{A} \dot{B}} \, dq_{\dot{A}} \underset{s}{\otimes} dq_{\dot{B}})
\end{equation}
The expansion 1-form which characterises the congruence of self-dual null strings is defined by \cite{biblio_2}
\begin{equation}
\label{forma_ekspansji}
\theta := \theta_{a} e^{a} = \phi^{4} (\Gamma_{423} e^{3} + \Gamma_{421} e^{1}) = \phi \, \frac{\partial \phi}{\partial p_{\dot{A}}} \, dq_{\dot{A}}
\end{equation}
Consequently, we can consider two cases
\begin{itemize}
\item if $\frac{\partial \phi}{\partial p_{\dot{A}}}=0 \rightarrow \theta=0$ and such a space is called \textsl{a nonexpanding $\mathcal{HH}$-space} (in this case one can put $\phi=1$)
\item if $\frac{\partial \phi}{\partial p_{\dot{A}}}\ne0 \rightarrow \theta \ne 0$ we have an  \textsl{expanding $\mathcal{HH}$-space}
\end{itemize}
When self-dual (or anti-self-dual) part of the Weyl spinor vanishes then we obtain the so called \textsl{heavenly space} ($\mathcal{H}$-\textsl{space}) with $\Lambda$. Theory of heavenly spaces was provided by many authors, but most of them do not include  cosmological constant $\Lambda$. Most of the papers on heavenly spaces theory take as a starting point nonexpanding congruence of the null strings. The geometry of heavenly spaces can be easily obtained from the geometry of nonexpanding hyperheavenly spaces (in the notation used in our work \cite{biblio_15} it is enough to set $F^{\dot{A}}=N^{\dot{A}}=n=\Lambda=0$ in all formulas).

However, as it has been shown in \cite{biblio_6}, if the cosmological constant $\Lambda \ne 0$, there do not exist nonexpanding congruences of the null strings. To obtain algebraic description of \textsl{heavenly spaces with $\Lambda$} we have to start from expanding hyperheavenly space. We do not consider here transition from expanding hyperheavenly spaces to heavenly spaces with $\Lambda$ with detailes. Simple analysis of the self-dual part of Weyl spinor proves, that $C_{ABCD}=0$ involves $\mu=\nu=0$ (compare \cite{biblio_4}, \cite{biblio_10}). Except this $\varkappa$ and $\gamma$ can be gauged to zero (significance of the functions $\mu$, $\nu$, $\varkappa$ and $\gamma$ can be found in \cite{biblio_2}, \cite{biblio_4}). With this simplifications made, one can present general structure of heavenly spaces with $\Lambda$.

\subsection{Geometry of $\mathcal{H}$-spaces with $\Lambda$ and the heavenly equation with $\Lambda$.}

The first step towards reducing the Einstein equations $R_{ab}=-\Lambda g_{ab}$, $\Lambda \ne 0$, consists in the observation that without any loss of generality one can put
\begin{equation}
\phi=J_{\dot{A}} \, p^{\dot{A}}  
\end{equation}
where $J_{\dot{A}}$ is a constant, nonzero spinor. By $K_{\dot{A}}$ we denote the inverse spinor defined by the relation
\begin{equation}
\label{connection_miedzy_KiJ}
K^{\dot{A}} J_{\dot{B}} - K_{\dot{B}} J^{\dot{A}} = \tau \, \delta^{\dot{A}}_{\dot{B}}
\ \ \ \ \textrm{where} \ \ \ \ \tau = K^{\dot{A}} J_{\dot{A}} \ne 0
\end{equation}
Then we obtain that there exists a holomorphic function $W$ (\textsl{the key function}) such that
\begin{subequations}
\begin{eqnarray}
\label{expanding_QAB}
Q^{\dot{A} \dot{B}} &=& -2 \, J^{(\dot{A}} \partial^{\dot{B})} W - \phi \, \partial^{\dot{A}} \partial^{\dot{B}} W+ \frac{\Lambda}{6\tau^{2}} K^{\dot{A}} K^{\dot{B}}
\\ 
\label{expanding_QAB_z_Omega}
&=& \partial^{(\dot{A}} \Omega^{\dot{B})} 
\end{eqnarray}
\end{subequations}
where
\begin{equation}
\label{postac_Omegi}
\Omega^{\dot{B}} 
= -\phi \, \partial^{\dot{B}} W - 3 J^{\dot{B}} \, W + \frac{\Lambda}{6\tau^{2}} K^{\dot{B}} \eta
= - \phi^{4} \, \partial^{\dot{B}} (\phi^{-3}W) + \frac{\Lambda}{6\tau^{2}} K^{\dot{B}} \eta
\end{equation}
and
\begin{equation}
\label{definicja_eta}
\eta := K^{\dot{A}} p_{\dot{A}} \ \ \ \ \ \rightarrow \ \ \ \ \ \tau  p^{\dot{A}} = \eta  \, J^{\dot{A}} + \phi \, K^{\dot{A}} 
\end{equation}
Finally, the Einstein equations can be reduced to a single nonlinear partial differential equation of the second order called the \textsl{heavenly equation with $\Lambda$}
\begin{equation}
\label{expanding_hyperheavenly_equation_form1}
\mathcal{T} +(\phi^{-1}J^{\dot{A}} \, W_{p^{\dot{A}}})^{2} + \phi^{-1} W_{p_{\dot{A}}q^{\dot{A}}}
-\frac{\Lambda}{3} \, \phi^{-2} \, \partial_{\phi} W = 0
\end{equation}
\begin{equation}
\label{form_of_T}
\mathcal{T} := \frac{1}{2}\phi^{-2} Q^{\dot{A} \dot{B}} Q_{\dot{A} \dot{B}} 
\end{equation}
\begin{equation}
\partial_{\phi} = \frac{1}{\tau} K^{\dot{A}} \partial_{\dot{A}} 
\ \ \ \ \ \ \ \ \partial_{\eta} = \frac{1}{\tau} J^{\dot{A}} \partial_{\dot{A}} 
\end{equation}
The form (\ref{expanding_hyperheavenly_equation_form1}) will be especially useful in calculations of Appendix  \ref{sekcja_szczegolowe_obliczenia}. Inserting into (\ref{expanding_hyperheavenly_equation_form1}) the explicit form of $\mathcal{T}$ given by (\ref{form_of_T}) with (\ref{expanding_QAB_z_Omega}) one gets the heavenly equation with $\Lambda$ in the following form
\begin{equation}
\label{expanding_hyperheavenly_equation_form2}
\frac{1}{2} \phi^{4} (\phi^{-2}W_{p_{\dot{B}}})_{p_{\dot{A}}}
                     (\phi^{-2}W_{p^{\dot{B}}})_{p^{\dot{A}}}
+\phi^{-1} W_{p_{\dot{A}} q^{\dot{A}}} 
- \frac{\Lambda}{6} \, \phi^{-1} \, W_{\phi \phi}
= 0
\end{equation}
where $W_{p^{\dot{A}}} \equiv \frac{\partial W}{\partial p^{\dot{A}}}$,  $W_{q_{\dot{A}}} \equiv \frac{\partial W}{\partial q_{\dot{A}}}, etc.$.
Multiplaying (\ref{expanding_hyperheavenly_equation_form2}) by $\phi$, one gets
\begin{equation}
\label{rownanie_HH_w_postaci_pochodnej}
0 = \partial^{\dot{A}} \bigg( \frac{1}{2} \phi W_{p^{\dot{A}}p^{\dot{B}}}W_{p_{\dot{B}}} - \frac{5}{2} J^{\dot{B}} W_{p^{\dot{B}}} W_{p^{\dot{A}}} + W_{q^{\dot{A}}} - \frac{\Lambda}{6 \tau} K_{\dot{A}} W_{\phi} \bigg)
\end{equation}
From (\ref{rownanie_HH_w_postaci_pochodnej}) we infer that there exists a holomorphic function $\Upsilon$ such that
\begin{equation}
\label{calka_pierwsza_row_HH}
 \frac{1}{2} \phi W_{p^{\dot{A}}p^{\dot{B}}}W_{p_{\dot{B}}} - \frac{5}{2} J^{\dot{B}} W_{p^{\dot{B}}} W_{p^{\dot{A}}} + W_{q^{\dot{A}}} - \frac{\Lambda}{6 \tau} K_{\dot{A}} W_{\phi}
= \Upsilon_{ p^{\dot{A}}}
\end{equation}
Then the connection 1-forms in spinorial formalism are
\begin{eqnarray}
\label{expanding_koneksja}
\mathbf{\Gamma}_{11}  &=& -\phi^{-1} J_{\dot{A}} \, e^{\dot{A}}                    
\\ \nonumber
\mathbf{\Gamma}_{12} &=& 
-\frac{1}{2} \phi \, \partial^{\dot{A}} (\phi \, Q_{\dot{A} \dot{B}}) \, e^{\dot{B}}
- \frac{3}{2} \phi^{-1} J_{\dot{B}} \, E^{\dot{B}}
\\ \nonumber
\mathbf{\Gamma}_{22} &=& -\phi^{2} \, (\eth^{\dot{A}} Q_{\dot{A}\dot{B}}) \, e^{\dot{B}} 
- \phi \, (Q_{\dot{A}\dot{B}}J^{\dot{A}}) E^{\dot{B}}
\\ \nonumber
\mathbf{\Gamma}_{\dot{A} \dot{B}} &=& 
\phi \Big( \phi \, \partial_{(\dot{A}} Q_{\dot{B})\dot{C}} + J^{\dot{D}} \in_{\dot{C} ( \dot{B}} Q_{\dot{A} ) \dot{D}} \Big)  e^{\dot{C}} + \phi^{-1} J_{(\dot{A}} E_{\dot{B})}
\end{eqnarray}
where
\begin{equation}
Q_{\dot{A}\dot{B}} J^{\dot{A}} =- \phi^{2} \, \left( J^{\dot{A}} (\phi^{-1} W_{p^{\dot{A}}})_{p^{\dot{B}}} + \frac{\Lambda}{6 \tau} \phi^{-2} K_{\dot{B}}  \right)
= \phi^{-2} \, \eth_{\dot{B}} \phi
\end{equation}
and 
\begin{equation}
\eth^{\dot{A}} Q_{\dot{A}\dot{B}} = \phi^{2}  J^{\dot{A}} W_{q^{\dot{A}}p^{\dot{B}}} 
\end{equation}
Decomposing the connection 1-forms according to
\begin{equation}
\label{rozklad_koneksji}
\mathbf{\Gamma}_{AB} = -\frac{1}{2} \, \mathbf{\Gamma}_{AB C \dot{D}} \, g^{C \dot{D}}
\ \ \ \ \ \ \ \ \  
\mathbf{\Gamma}_{\dot{A} \dot{B}} = -\frac{1}{2} \, \mathbf{\Gamma}_{\dot{A} \dot{B} C \dot{D}} \, g^{C \dot{D}}
\end{equation}
we get
\begin{eqnarray}
&&\mathbf{\Gamma}_{111 \dot{D}} =0  \ \ \ \ \ \ \ \ \ \ \ \ \ \ \ \ \  \ \ \ \ \ \ \ \ \ \ \ \ \ 
\mathbf{\Gamma}_{112 \dot{D}} = -\sqrt{2} \, \phi^{-1} \, J_{\dot{D}}
\\ \nonumber
&&\mathbf{\Gamma}_{121 \dot{D}} = \frac{3}{\sqrt{2}} \phi^{-1} \, J_{\dot{D}} \ \ \ \ \ \ \ \ \ \ \ \ \ \ \ \ \ \ 
  \mathbf{\Gamma}_{122 \dot{D}} = - \frac{1}{\sqrt{2}} \, \phi \, \partial^{\dot{A}} (\phi \, Q_{\dot{A}\dot{D}})
\\ \nonumber
&& \mathbf{\Gamma}_{221 \dot{D}} = \sqrt{2} \phi \, Q_{\dot{D}\dot{A}} J^{\dot{A}}
\ \ \ \ \ \ \ \ \ \ \ \ \ \ \ \,
\mathbf{\Gamma}_{222 \dot{D}} = -\sqrt{2} \phi^{2} \, \eth^{\dot{A}} Q_{\dot{A}\dot{D}}
\\ \nonumber
&&\mathbf{\Gamma}_{\dot{A} \dot{B} 1 \dot{D}} = -\sqrt{2} \phi^{-1} \, J_{(\dot{A}} \in_{\dot{B})\dot{D}}
\ \ \ \ \ \ 
\mathbf{\Gamma}_{\dot{A} \dot{B} 2 \dot{D}} = - \sqrt{2} \, \phi \Big(  
\phi \, \partial_{(\dot{A}} Q_{\dot{B})\dot{D}} + J^{\dot{C}} \in_{\dot{D}(\dot{B}} Q_{\dot{A})\dot{C}} \Big) 
\end{eqnarray}
The conformal curvature is given by
\begin{eqnarray}
\label{expanding_krzywizna}
C_{ABCD} &=& 0
\\ \nonumber
C_{\dot{A}\dot{B}\dot{C}\dot{D}} &= &
\phi^{3} \, W_{p^{\dot{A}}p^{\dot{B}}p^{\dot{C}}p^{\dot{D}}}
\end{eqnarray}
Substituting (\ref{expanding_QAB}) into the metric (\ref{metryka_ogolna}) one finds
\begin{eqnarray}
\label{metrykaprzestrzeni_HH_z_ekspanja}
ds^2 &=& \phi^{-2} \bigg\{ 2\tau^{-1} (d \eta \underset{s}{\otimes} d w - d \phi \underset{s}{\otimes} dt) + 2 \, \Big( - \phi \, W_{\eta\eta} +\frac{\Lambda}{6\tau^{2}}  \Big) 
dt \underset{s}{\otimes}dt  
\\ \nonumber
&& \ \ \ \ \ \ \ \ \ \ +4 \left( - \phi \, W_{\eta \phi} +  \, W_{\eta}\right) dw \underset{s}{\otimes}dt
  + 2 \left(  - \phi \, W_{\phi \phi} + 2 \, W_{\phi}         \right) dw \underset{s}{\otimes}dw \bigg\} \ \ \ \ \ \ 
\end{eqnarray}
where
\begin{equation}
t := K^{\dot{A}}q_{\dot{A}}
\ \ \ \ \ \ \ \ \ \ \ \
w := J_{\dot{A}}q^{\dot{A}}
\ \ \ \ \ \rightarrow \ \ \ \ \ 
q_{\dot{B}} = \frac{1}{\tau} \, (t J_{\dot{B}} + w K_{\dot{B}})
\end{equation}
In terms of the coordinates $(\phi, \eta, w,t)$ the heavenly equation with $\Lambda$ (\ref{expanding_hyperheavenly_equation_form2}) reads
\begin{eqnarray}
\label{rownanie_niebianskie_zLambda_postac_ostateczna}
&&  W_{\eta \eta}W_{\phi \phi} - W_{\eta \phi}W_{\eta \phi} + 2 \phi^{-1} W_{\eta} W_{\eta \phi} - 2 \phi^{-1} W_{\phi}W_{\eta\eta}  
\ \ \ \ \ \ 
\\ \nonumber
&& + (\tau \phi)^{-1} \big( W_{w\eta}-W_{t\phi} \big) - \frac{\Lambda}{6 \tau^2} \phi^{-1} W_{\phi \phi} = 0
\end{eqnarray}
and Eq. (\ref{calka_pierwsza_row_HH}) takes the form
\begin{subequations}
\begin{eqnarray}
\label{rozpisana_calka_pierwsza_1}
&& \Upsilon_{\phi} = \frac{\tau}{2} \phi \Big( W_{\phi \phi} W_{\eta} - W_{\eta \phi} W_{\phi} \Big) - \frac{5 \tau}{2} \, W_{\eta} W_{\phi} + W_{w}
\\ 
\label{rozpisana_calka_pierwsza_2}
&& \Upsilon_{\eta} = \frac{\tau}{2} \phi \Big( W_{\eta \phi} W_{\eta} - W_{\eta \eta} W_{\phi} \Big) - \frac{5 \tau}{2} \, W_{\eta} W_{\eta} + W_{t} + \frac{\Lambda}{6 \tau} \, W_{\phi}
\end{eqnarray}
\end{subequations}

\subsection{Gauge freedom.}
\label{subsection_gauge}

The problem of finding the coordinate freedom in expanding $\mathcal{HH}$-spaces with $\Lambda$ and, consequently, in $\mathcal{H}$-spaces with $\Lambda$ is similiar to that problem in nonexpanding $\mathcal{HH}$-spaces (compare \cite{biblio_2, biblio_10, biblio_16}). The form of metric (\ref{metryka_ogolna}) admitts the coordinate gauge freedom
\begin{eqnarray}
\label{gaug_wspolrzednych}
q'_{\dot{A}} &=& q'_{\dot{A}} (q_{\dot{B}})
\\ \nonumber
p'^{\dot{A}} &=& \lambda^{-1} \, D^{-1 \ \; \dot{A}}_{\ \ \; \dot{B}}  \, p^{\dot{B}} + \sigma^{\dot{A}} 
\ \ \ \ , \ \ \ \ 
\lambda=\lambda(q_{\dot{A}}), \ \ \sigma^{\dot{A}}=\sigma^{\dot{A}} (q_{\dot{B}})
\end{eqnarray}
The functions $\phi$ and $Q^{\dot{A}\dot{B}}$ transform under (\ref{gaug_wspolrzednych}) as follows
\begin{equation}
\phi'^{-2} = \lambda \, \phi^{-2} \ \ \ \ , \ \ \ \ 
Q'^{\dot{A}\dot{B}} = \lambda^{-1} \, D^{-1 \ \; \dot{A}}_{\ \ \; \dot{C}}  \,
    D^{-1 \ \; \dot{B}}_{\ \ \; \dot{D}} \, Q^{\dot{C}\dot{D}} + 
   D^{-1 \ \; (\dot{A}}_{\ \ \; \dot{R}} \, 
    \frac{\partial p'^{\dot{B})}}{\partial q_{\dot{C}}} 
\end{equation}
where we put
\begin{eqnarray}
D_{\dot{A}}^{\ \ \dot{B}} 
&:=& \frac{\partial q'_{\dot{A}}}{\partial q_{\dot{B}}} 
= \Delta \, \frac{\partial q^{\dot{B}}}{\partial q'^{\dot{A}}}
= \lambda \Delta \, \frac{\partial p'_{\dot{A}}}{\partial p_{\dot{B}}}
= \lambda^{-1} \frac{\partial p^{\dot{B}}}{\partial p'^{\dot{A}}}
\\ \nonumber
D^{-1 \ \; \dot{B}}_{\ \ \; \dot{A}} 
&:=& \frac{\partial q_{\dot{A}}}{\partial q'_{\dot{B}}}
= \Delta^{-1} \, \frac{\partial q'^{\dot{B}}}{\partial q^{\dot{A}}}
= \lambda \, \frac{\partial p'^{\dot{B}}}{\partial p^{\dot{A}}}
= \lambda^{-1} \Delta^{-1} \, \frac{\partial p_{\dot{A}}}{\partial p'_{\dot{B}}}
\end{eqnarray}
and $\Delta$ is the following determinant
\begin{equation}
\Delta := \det 
\left( \frac{\partial q'_{\dot{A}}}{\partial q_{\dot{B}}} \right)
 = \frac{1}{2} \, D_{\dot{A}\dot{B}} D^{\dot{A}\dot{B}}
\end{equation}
Symbol $D^{\dot{A}}_{\ \, \dot{B}}$ is essentially different of $D_{\dot{A}}^{\ \ \dot{B}}$. Note, that
\begin{equation}
 D^{\dot{A}}_{\ \, \dot{B}} = - \Delta \, D^{-1 \ \; \dot{A}}_{\ \ \; \dot{B}}  \ \ \ \ \ \ \ \ \ \ 
D^{-1\, \dot{A}}_{\ \ \ \ \; \dot{B}} = - \frac{1}{\Delta} \, D_{\dot{B}}^{\ \ \dot{A}} 
\end{equation}
It is easy to see, that the transformation (\ref{gaug_wspolrzednych}) is equivalent to a spinorial transformation of the tetrad $g^{A\dot{B}}$ given by (\ref{tetrada_spinorowa}), defined by the matrices 
\begin{eqnarray}
\label{transformacja_spinorowa}
L^{A}_{\ \; B} &=& \left[ \begin{array}{cc}
       \lambda^{-1}\Delta^{-\frac{1}{2}} & h\phi^{2}  \Delta^{\frac{1}{2}}   \\
       0 &  \lambda \Delta^{\frac{1}{2}}  
       \end{array} \right]
\\ \nonumber
M^{\dot{A}}_{\ \; \dot{B}} &=& \Delta^{\frac{1}{2}} \, D^{-1 \ \ \dot{A}}_{\ \ \; \dot{B}}
\end{eqnarray}
where
\begin{equation}
2h := - \lambda^{-1} \, D^{-1 \ \ \dot{R}}_{\ \ \; \dot{S}} \, 
\frac{\partial p^{\dot{S}}}{\partial q'^{\dot{R}}} 
= \frac{1}{\Delta} \, D_{\dot{S}}^{\ \ \dot{R}} \, 
\frac{\partial p'^{\dot{S}}}{\partial q^{\dot{R}}} 
= \frac{\partial \sigma^{\dot{R}}}{\partial q'^{\dot{R}}} + \frac{1}{\Delta} \frac{\partial (\lambda^{-1})}{\partial q^{\dot{R}}} p^{\dot{R}}
\end{equation}
It is convenient to assume that the gauge transformations maintain $J'^{\dot{A}} = J^{\dot{A}}=\textrm{const}$ and $K'^{\dot{A}} = K^{\dot{A}}=\textrm{const}$. This obviously  puts some restrictions on the gauge freedom. After some analysis one finds that 
\begin{equation}
\label{jawna_postac_gaugu}
w'=w'(w) \ \ \ , \ \ \ t'=t'(w,t) \ \ \ , \ \ \ t'_{t} = \lambda^{-\frac{1}{2}} \ \ \ , \ \ \ \Delta = w'_{w} \lambda^{-\frac{1}{2}} \ \ \ , \ \ \ \sigma^{\dot{A}} = \sigma J^{\dot{A}}
\end{equation}
\begin{equation}
\nonumber
D_{\dot{A}}^{\ \ \dot{B}} =  - \frac{1}{\tau} \, (t'_{w} J_{\dot{A}} + w'_{w} K_{\dot{A}} ) J^{\dot{B}} 
+ \frac{1}{\tau} \, \lambda^{-\frac{1}{2}} J_{\dot{A}}  K^{\dot{B}}
\end{equation}
where $\sigma=\sigma(w,t)$ is an arbitrary function.
\newline
Transformation law for $\eta$ reads
\begin{equation}
\eta' = \frac{1}{\lambda w'_{w}} \, \eta + \lambda^{-\frac{1}{2}} \frac{t'_{w}}{w'_{w}} \, \phi + \tau \sigma
\end{equation}
Straightforward but long calculations show that the function $W$ transforms under (\ref{jawna_postac_gaugu}) as follows \cite{biblio_10}
\begin{eqnarray}
\label{transformacja_funkcji_kluczowej}
(w'_{w})^{2} \lambda^{\frac{3}{2}} W' &=& W  - \frac{1}{3} L \, \phi^{3}
- \frac{1}{2\tau} \lambda^{\frac{1}{2}} w'_{w} \, 
\frac{\partial}{\partial q_{(\dot{A}}} 
\bigg( \frac{1}{w'_{w}} \frac{\partial t'}{\partial q_{\dot{B})}} \bigg) \, p_{\dot{A}} p_{\dot{B}} 
\\ \nonumber
&&+ \bigg(  -\frac{1}{2} \lambda w'_{w} \, \frac{\partial \sigma}{\partial q_{\dot{A}}}
+\frac{\Lambda}{12 \tau^{2}} \, 
(\lambda {t'_{w}}^{2} \, J^{\dot{A}} - 2 \lambda^{\frac{1}{2}} t'_{w} \, K^{\dot{A}}) \bigg) \,  p_{\dot{A}} - M
\end{eqnarray}
where $M$ is an arbitrary function of $q_{\dot{A}}$ only and $L=L(w)$ is function of $w$, satisfying the relation
\begin{equation}
0=  - (w'_{w})^{\frac{1}{2}} [(w'_{w})^{-\frac{1}{2}}]_{ww} + \frac{\Lambda}{3} L
\end{equation}

\setlength\arraycolsep{2pt}
\setcounter{equation}{0}

\section{General properties of Killing vectors.}
\label{sekcja_Killing}

\subsection{Killing equations and their integrability conditions in spinorial formalism.}

Define the spin-tensor $g^{aA \dot{B}}$ by the relation $g^{A \dot{B}} = g_{a}^{\ A \dot{B}} \, e^{a}$. Hence, $-\frac{1}{2} g^{aA\dot{B}}g_{bA\dot{B}} = \delta^{a}_{b}$ and $-\frac{1}{2}g^{aA\dot{B}}g_{aC\dot{D}} = \delta^{A}_{C} \delta^{\dot{B}}_{\dot{D}}$. The operators $\partial^{A\dot{B}}$ and $\nabla^{A\dot{B}}$ are the spinorial images of operators $\partial^{a}$ and $\nabla^{a}$, respectively, given by
\begin{equation}
\partial^{A\dot{B}} = g_{a}^{\  A \dot{B}} \partial^{a}  \ \ \ \ \ \ \ \ \ \ \ \ \ \ 
\nabla^{A\dot{B}} = g_{a}^{\  A \dot{B}} \nabla^{a} 
\end{equation}
In the basis (\ref{baza_dualna})
\begin{equation}
\label{rozklad_operatora}
\partial^{A\dot{B}} = \sqrt{2} \, (\delta^{A}_{1} \, \eth^{\dot{B}} - \delta^{A}_{2} \, \partial^{\dot{B}}) = \sqrt{2} \, \left[ \begin{array}{cc}
                           \ \; \eth^{\dot{1}} & \ \; \eth^{\dot{2}}   \\
                           -\partial^{\dot{1}} & -\partial^{\dot{2}}  
                            \end{array} \right]
\end{equation}
A conformal Killing vector $K$ can be written as
\begin{equation}
\label{ogolna_postac_wektora_Killll}
K = K^{a} \, \partial_{a} = -\frac{1}{2} K_{A\dot{B}} \partial^{A\dot{B}} = k_{\dot{B}} \eth^{\dot{B}} - h_{\dot{B}} \partial^{\dot{B}}
\end{equation}
where we use the decomposition
\begin{equation}
\label{rozklad_wektora_Killinga}
K_{A\dot{B}} = - \sqrt{2} \, (\delta^{1}_{A} \,k_{\dot{B}} + \delta^{2}_{A} \,h_{\dot{B}} ) \,  
= - \sqrt{2} \,  \left[ \begin{array}{cc}
                            k_{\dot{1}} & k_{\dot{2}}   \\
                           h_{\dot{1}} & k_{\dot{2}}  
                            \end{array} \right]
\end{equation}
Components $K^{a}$ and $K_{A \dot{B}}$ are related by 
\begin{equation}
K^{a} = - \frac{1}{2} g^{a A \dot{B}} \, K_{A \dot{B}} \ \ \leftrightarrow \ \ 
K_{A \dot{B}} = g_{a  A \dot{B}} \, K^{a}
\end{equation}
\textsl{Conformal Killing equations with conformal factor $\chi$} read
\begin{equation}
\nabla_{(a} K_{b)} = \chi \, g_{ab}
\end{equation}
or in spinorial form
\begin{equation}
\label{wyjsciowa_postac_ro_Killinga}
\nabla_{A}^{\ \; \dot{B}} K_{C}^{\ \dot{D}} + \nabla_{C}^{\ \; \dot{D}} K_{A}^{\ \dot{B}} = -4 \chi \in_{AC} \in^{\dot{B}\dot{D}}
\end{equation}
which is equivalent to the following system of equations
\begin{subequations}
\begin{eqnarray}
\label{rozklad_rownania_Killinga}
E_{AC}^{\ \ \ \dot{B}\dot{D}} &\equiv& \nabla_{(A}^{\ \; \dot{(B}} K_{C)}^{\ \; \dot{D)}} = 0 
\\
\label{definicja_chi}
E &\equiv& \nabla^{N\dot{N}} K_{N\dot{N}} +8 \chi = 0
\end{eqnarray}
\end{subequations}
From (\ref{rozklad_rownania_Killinga}) and (\ref{definicja_chi}) it follows that
\begin{equation}
\label{ogolne_rownanie_Killinga_spinnorowo}
\nabla_{A}^{\ \; \dot{B}} K_{C}^{\ \; \dot{D}} = l_{AC} \in^{\dot{B}\dot{D}} + l^{\dot{B}\dot{D}} \in_{AC} - 2 \chi  \in_{AC} \in^{\dot{B}\dot{D}}
\end{equation}
with
\begin{equation}
\label{definicja_eli}
l_{AC} := \frac{1}{2} \nabla_{(A}^{\ \ \dot{N}} K_{C)\dot{N}} \ \ \ \ \ \ \ \ \ \ \ \ 
l^{\dot{B}\dot{D}} := \frac{1}{2} \nabla^{N(\dot{B}} K_{N}^{\ \; \dot{D})}
\end{equation}
In \cite{biblio_17} the integrability conditions of (\ref{rozklad_rownania_Killinga}) and (\ref{definicja_chi}) have been found. For the self-dual Einstein space ($C_{AB\dot{C}\dot{D}}=0=C_{ABCD}$, $R=-4 \Lambda$) these conditions consist of the following equations
\begin{subequations}
\begin{eqnarray}
\label{integrability_c_L}
L_{RST}^{\ \ \ \ \; \dot{A}} &\equiv& \nabla_{R}^{\ \; \dot{A}} l_{ST} + \frac{2}{3} \Lambda \in_{R(S}K_{T)}^{\ \ \dot{A}} + 2 \in_{R(S} \nabla_{T)}^{\ \ \dot{A}} \chi =0 \ \ \ \ \ \ \ \ 
\\
L_{\dot{R}\dot{S}\dot{T}}^{\ \ \ \ \; A} &\equiv& \nabla^{A}_{\ \dot{R}} l_{\dot{S}\dot{T}} + \frac{2}{3} \Lambda \, \in_{\dot{R} ( \dot{S}}K^{A}_{\ \ \dot{T})} + 2 \in_{\dot{R} ( \dot{S}} \nabla^{A}_{\ \ \dot{T})} \chi=0
\\
\label{integrability_c_N}
N_{AB}^{\ \ \ \dot{A}\dot{B}} &\equiv& \nabla_{A}^{\ \; \dot{A}} \nabla_{B}^{\ \; \dot{B}} \chi - \frac{2}{3} \Lambda \chi \in_{AB} \in^{\dot{A}\dot{B}} =0
\\
\label{integrability_c_R}
R_{\dot{A}\dot{B}\dot{C}}^{\ \ \ \ \ A} &\equiv& C^{\dot{N}}_{\ \; \dot{A}\dot{B}\dot{C}} \nabla^{A}_{\ \; \dot{N}} \chi = 0
\end{eqnarray}
\end{subequations}
Structure of the heavenly spaces with $\Lambda$ puts some strong restrictions on the conformal factor $\chi$. Namely, it can be proved that if $C_{\dot{A}\dot{B}\dot{C}\dot{D}} \ne 0$ then $\chi=0$ \cite{biblio_18}. Indeed, from (\ref{integrability_c_R}) we conclude, that $\nabla_{A}^{\ \; \dot{A}} \chi$ is the quadruple Debever-Penrose spinor and consequently, the conformal symmetries can eventually appear only in the heavenly space with $\Lambda$ of the type $[-] \otimes [ \textrm{N} ]$. However, as is well known, two quadruple DP spinors are necessarily lineary dependent so $\nabla^{1 \dot{A}} \chi$ has to be proportional to $\nabla^{2 \dot{A}} \chi$ or, equivalently
\begin{equation}
\label{one_quadruple_DP_spinor_condition}
\nabla_{A\dot{A}} \chi  \cdot   \nabla^{A\dot{A}} \chi =0
\end{equation}
Acting on (\ref{one_quadruple_DP_spinor_condition}) with $\nabla_{B}^{\ \; \dot{B}}$ and using (\ref{integrability_c_N}) one quickly obtains
\begin{equation}
\Lambda \chi \, \nabla_{B}^{\ \; \dot{B}} \chi =0
\end{equation}
Hence if $\Lambda \ne 0$ then $\nabla_{B}^{\ \; \dot{B}} \chi =0$. Finally, using (\ref{integrability_c_N})) we get $\chi =0$. 

Gathering: there is no conformal or homothetic symmetries in the spaces of the type $[-] \otimes [\textrm{I,II,III,D,N}]$ with $\Lambda \ne 0$. Conformal or homothetic symmetries with $\Lambda \ne 0$ can exist only in the space of the type $[-] \otimes [-]$  i.e., in the de-Sitter spacetime. We do not consider these spaces here.

Consequently, in what follows we consider $\chi=0$ and so we deal with isometries and Killing vectors.

\subsection{Master equation.}

Assuming $\chi=0$ we are left with the integrability conditions (\ref{integrability_c_L}) with $\chi=0$. Careful analysis of these conditions lead to the conclusion that the components of Killing vector (\ref{ogolna_postac_wektora_Killll}) are
\begin{subequations}
\begin{eqnarray}
\label{postac_k}
k^{\dot{B}} &=& \phi^{-2} (\alpha  p^{\dot{B}} + \delta^{\dot{B}})
\\ 
\label{postac_h}
h^{\dot{B}} &=& \phi^{2} \, k_{\dot{S}} \, Q^{\dot{S}\dot{B}} -2 \alpha \, \Omega^{\dot{B}} - \frac{\partial \alpha}{\partial q^{\dot{S}}} \, p^{\dot{S}} p^{\dot{B}} 
+ \frac{\partial \delta_{\dot{S}}}{\partial q_{\dot{B}}} \, p^{\dot{S}} + \epsilon \, J^{\dot{B}}
+  \frac{2}{\tau} K_{\dot{S}} J_{\dot{N}} \frac{\partial \delta^{\dot{S}}}{\partial q_{\dot{N}}} \, p^{\dot{B}}
 \ \ \ \ \ \ \ \  
\end{eqnarray}
\end{subequations}
where $Q^{\dot{A}\dot{B}}$ is given by (\ref{expanding_QAB}), $\Omega^{\dot{B}}$ by (\ref{postac_Omegi}), $\epsilon=\epsilon(q^{\dot{M}})$ and $\alpha=\alpha(q^{\dot{M}})$ are arbitrary functions.
\newline
In the base $\Big( \frac{\partial}{\partial q^{\dot{A}}}, \frac{\partial}{\partial p^{\dot{B}}} \Big)$ the Killing vector has the form
\begin{eqnarray}
\label{ostateczna_postac_wektora_Kili}
K&=& (\alpha  p^{\dot{B}} + \delta^{\dot{B}}) \frac{\partial}{\partial q^{\dot{B}}} 
\\ \nonumber
&&- \bigg[  
-2 \alpha \, \Omega^{\dot{B}} - \frac{\partial \alpha}{\partial q^{\dot{S}}} \, p^{\dot{S}} p^{\dot{B}} 
+ \frac{\partial \delta_{\dot{S}}}{\partial q_{\dot{B}}} \, p^{\dot{S}} + \epsilon \, J^{\dot{B}}
+  \frac{2}{\tau} K_{\dot{S}} J_{\dot{N}} \frac{\partial \delta^{\dot{S}}}{\partial q_{\dot{N}}} \, p^{\dot{B}} \bigg] \frac{\partial}{\partial p^{\dot{B}}}
\end{eqnarray}
The system of ten Killing equations can be reduced to one, \textsl{master equation}
\begin{eqnarray}
\label{expanding_master_equation}
\pounds_{K} W &=& 4 \alpha \Upsilon -W \bigg( -3 \, \frac{\partial \alpha}{\partial q^{\dot{S}}} \, p^{\dot{S}}  + \frac{1}{\tau}  \, \frac{\partial \delta^{\dot{S}}}{\partial q_{\dot{N}}} \, (3K_{\dot{S}} J_{\dot{N}} + 2 K_{\dot{N}} J_{\dot{S}} )\bigg) 
\\ \nonumber
&& - 4 \phi^{3} \xi + \phi^{-1} \mathcal{P}
\end{eqnarray}
where $\pounds_{K} W=KW$ is the Lie derivative of the key function, $\xi=\xi(q^{\dot{M}})$ is an arbitrary function and $\mathcal{P}$ is a fourth order polynomial in $p^{\dot{S}}$
\begin{eqnarray}
\label{definicja_wielomianu_P}
\mathcal{P} &:=& - \frac{1}{2 \tau^{3}} \frac{\partial^{2} \alpha}{\partial q^{\dot{S}} \partial q^{\dot{R}}} \, \phi \eta \, \Big( \phi^{2} K^{\dot{S}} K^{\dot{R}} + \phi \eta K^{\dot{S}}J^{\dot{R}} + \frac{1}{3} \eta^{2} J^{\dot{R}} J^{\dot{S}} \Big) 
\\ \nonumber
&&+ \frac{1}{2} \frac{\partial^{2} \delta^{\dot{S}}}{\partial q_{\dot{B}} \partial q_{\dot{R}}} \, p_{\dot{R}} p_{\dot{S}} p_{\dot{B}}   + \frac{\Lambda}{6 \tau^{2}} \frac{\partial \alpha}{\partial q^{\dot{S}}} \, p^{\dot{S}} \eta^{2} + \frac{1}{2} \frac{\partial \epsilon}{\partial q_{\dot{S}}} \, \phi p_{\dot{S}} 
\\ \nonumber
&&+ \frac{\Lambda}{6 \tau^{3}} \, K_{\dot{S}} K_{\dot{N}} \, \frac{\partial \delta^{\dot{S}}}{\partial q_{\dot{N}}} \, \eta \phi  + \frac{Y}{3 \tau} \, \phi
\end{eqnarray}
where $Y=Y(q^{\dot{M}})$ is an arbitrary function of $q^{\dot{M}}$.
\newline
Then we are left with two integrability conditions which must be fullfilled
\begin{subequations}
\begin{eqnarray}
\label{integrability_condition_1}
&& J_{\dot{A}} J_{\dot{B}} \, \frac{\partial \delta^{\dot{A}}}{\partial q_{\dot{B}}} +\frac{\Lambda \alpha}{3}  = 0
\\ 
\label{integrability_condition_3}
&&8 \Lambda    \xi
+ \frac{1}{\tau^{3}}  \, J_{\dot{N}} K_{\dot{S}} K_{\dot{R}} K_{\dot{Z}} \, \frac{\partial^{3} \delta^{\dot{N}}}{\partial q_{\dot{S}} \partial q_{\dot{R}} \partial q_{\dot{Z}}} =0
\end{eqnarray}
\end{subequations}
From (\ref{definicja_eli}) with (\ref{ostateczna_postac_wektora_Kili}) by long and tedious calculations we get $l_{AB}$
\begin{subequations}
\begin{eqnarray}
\label{postac_l11}
l_{11} &=& -2\phi^{-3} \, J_{\dot{N}}\delta^{\dot{N}}
\\
\label{postac_l12}
l_{12} &=& 2\phi^{-1} \, J^{\dot{N}} \delta_{\dot{N}} \, J_{\dot{B}} \partial^{\dot{B}} W - \phi^{-1} \, K^{\dot{N}}\delta_{\dot{N}} \, \frac{\Lambda}{3\tau} + \frac{1}{\tau} \, J_{\dot{N}} K_{\dot{A}} \, \frac{\partial \delta^{\dot{N}}}{\partial q_{\dot{A}}}
\\ \nonumber
&&+ \frac{1}{\tau} \, \frac{\eta}{\phi} \, J_{\dot{N}} J_{\dot{A}} \, \frac{\partial \delta^{\dot{N}}}{\partial q_{\dot{A}}} 
\\
\label{postac_l22}
l_{22} &=& 2\phi J_{\dot{N}}\delta^{\dot{N}} \bigg( \frac{\Lambda}{3}  \, W_{\phi} -  (J_{\dot{S}}\partial^{\dot{S}}W)^{2}   \bigg) -2\phi^{2} J^{\dot{S}} \, \frac{\partial \delta_{\dot{S}}}{\partial q^{\dot{N}}} \, \partial^{\dot{N}}W  
\\ \nonumber
&& -2 \Lambda \alpha  \phi \, W  -\phi  p_{\dot{S}}p_{\dot{A}}J_{\dot{N}} \frac{\partial^{2} \delta^{\dot{N}}}{\partial q_{\dot{A}} \partial q_{\dot{S}}}
\\ \nonumber
&& - \frac{\Lambda}{3 \tau^{2}} \phi \frac{\partial \delta^{\dot{S}}}{\partial q_{\dot{N}}} \Big( 2\phi \, K_{\dot{N}}K_{\dot{S}} + \eta \, (2K_{\dot{S}}J_{\dot{N}}+K_{\dot{N}}J_{\dot{S}}) \Big) 
- \frac{\Lambda \epsilon}{3} \, \phi
\end{eqnarray}
\end{subequations}

\subsection{Transformation rules.}

Transformation rules for $\alpha$ and $\delta^{\dot{A}}$ read
\begin{subequations}
\begin{eqnarray}
\alpha' &=& w'_{w} \lambda^{\frac{1}{2}} \alpha
\\ 
\label{transformacja_delty}
\delta'^{\dot{A}} &=& w'_{w} \lambda^{-\frac{1}{2}} D^{-1 \ \ \dot{A}}_{\ \ \; \dot{B}} \delta^{\dot{B}} 
   - \alpha w'_{w} \lambda^{\frac{1}{2}} \sigma J^{\dot{A}}
\end{eqnarray}
\end{subequations}
Using the decomposition
\begin{equation}
\label{rozklad_delty_na_aib}
\tau \delta^{\dot{A}} = a  K^{\dot{A}} + b J^{\dot{A}} \ \ \ \leftrightarrow \ \ \ a:= J_{\dot{A}} \delta^{\dot{A}} \ \ \ , \ \ \ b:= -K_{\dot{A}} \delta^{\dot{A}}
\end{equation}
where $b=b(w,t)$, $a=a(w,t)$, one can rewrite (\ref{transformacja_delty}) in the form
\begin{subequations}
\begin{eqnarray}
\label{transformacja_a}
a' &=& w'_{w} a
\\
\label{transformacja_b}
b' &=& \lambda^{-\frac{1}{2}} b + t'_{w} a - \tau \sigma \alpha \lambda^{\frac{1}{2}} w'_{w}
\end{eqnarray}
\end{subequations}
then $\epsilon$ transforms as 
\begin{eqnarray}
\label{transformacja_epsilon}
\epsilon' &=& (\lambda w'_{w})^{-1} \epsilon +\frac{6 \alpha M}{\lambda w'_{w}} - \tau \lambda^{\frac{1}{2}}  (\sigma^{2} \alpha')_{t} + 3\lambda^{\frac{1}{2}} t'_{w} \sigma a_{t}
\\ \nonumber
&&- \sigma \Big[ a_{w} -2b_{t} + a \, ( \ln \sigma \lambda w'_{w})_{w} + b \, (\ln \sigma \lambda)_{t} \Big]
\end{eqnarray}
From the invariancy of the master equation (\ref{expanding_master_equation}) one can find the transformation laws for $\xi$ and $Y$. Under the assumption that $\alpha=0$ we get

\begin{eqnarray}
\label{transformacja_alpha}
(w'_{w})^{2} 4\xi' &=& 4 \xi + \frac{1}{3} a L_{w}  + \frac{2}{3} L  a_{w}
\\
\label{transformacja_beta}
(w'_{w})^{2} \lambda^{\frac{3}{2}} \frac{Y'}{3 \tau} &=& \frac{Y}{3 \tau} -\frac{1}{2} \tau \sigma \lambda^{2} (w'_{w})^{2} \frac{\partial \epsilon'}{\partial t} + \frac{1}{2} \tau (\sigma \lambda w'_{w})^{2} \partial_{t} \Big( \lambda^{\frac{1}{2}} \,b'_{t} \Big)
\\ \nonumber
&& +\frac{1}{2}\tau \lambda \epsilon \sigma_{t} \, w'_{w} +M ( 3 b_{t} -2 a_{w} ) - a M_{w} - b M_{t} 
\\ \nonumber
&& +\frac{\Lambda}{6 \tau} \epsilon \lambda^{\frac{1}{2}} t'_{w} + \frac{\Lambda}{6 \tau}\sigma w'_{w} \lambda^{\frac{3}{2}} \Big(\partial_{w} - \frac{t'_{w}}{\lambda^{-\frac{1}{2}}} \, \partial_{t} \Big) b' 
\end{eqnarray}

\renewcommand{\arraystretch}{1.5}
\setlength\arraycolsep{2pt}
\setcounter{equation}{0}

\section{The classification of $\mathcal{H}$-spaces with $\Lambda$ admitting Killing symmetries.}

\subsection{Preparatory analysis.}

In this section we give the classification of all $\mathcal{H}$-spaces with $\Lambda$ admitting Killing symmetries. We assume, that $\Lambda \ne 0$, $\chi=0$ and $C_{ABCD}=0$. First, we prove some lemmas which are crucial in our further analysis.
\newline
\newline
\textbf{Lemma 4.1}
\newline
If $K_{A \dot{B}}$ is a Killing vector then the spinor $l_{AB}$ defined by (\ref{ogolne_rownanie_Killinga_spinnorowo}) and (\ref{definicja_eli}) is nonzero.
\newline
\newline
\textbf{Proof}
\newline
Assume that $l_{AB}=0$. Then, from (\ref{integrability_c_L}) with $\Lambda \ne 0$ and $\chi=0$ one concludes that $K_{A \dot{B}}=0$. Thus, we arrive at the contradiction and this ends the proof. $\blacksquare$
\newline
\newline
The next lemma is fundamental in the cases of null Killing vectors
\newline
\newline
\textbf{Lemma 4.2}
\newline
Let $K_{A \dot{B}}$ be a Killing vector. Then the following statements are equivalent
\newline
$(i)$ $K_{A \dot{B}}$ is a null vector
\newline
$(ii)$ $K_{A \dot{B}}$ is of the form $K_{A \dot{B}}=m_{A} n_{\dot{B}}$
\newline
$(iii)$ $l_{AB}$ is of the form $l_{AB}=\mu \, m_{A} m_{B}$, where $\mu$ is some nowhere vanishing function
\newline
$(iv)$ $l_{AB}l^{AB} = 2 \det (l_{AB}) =0$
\newline
\newline
\textbf{Proof}
The proofs of equivalences $(i) \Leftrightarrow (ii)$ and $(iii) \Leftrightarrow (iv)$ are trivial. Therefore, consider the case $(ii) \Rightarrow (iii)$. Inserting $K_{C \dot{D}} = m_{C} n_{\dot{D}}$ and $K_{A \dot{B}}=m_{A} n_{\dot{B}}$ into the Killing equations (\ref{wyjsciowa_postac_ro_Killinga}) and multiplying both sides by $m^{A}m^{C}$ one gets
\begin{equation}
\label{rownanie_struny_ekspandujacej}
m^{A} m^{C} \nabla_{A \dot{B}} m_{C}=0
\end{equation}
This means that the spinor $m_{A}$ defines a congruence of self-dual null strings in the sense that the 2-dimensional holomorphic distribution $\{m_{A}t_{\dot{B}}, m_{A}u_{\dot{B}} \}$, $t_{\dot{B}} u^{\dot{B}} \ne 0$ is integrable and its integrable manifolds constitute the congruence of self-dual null strings. Note that multiplying the Killing equations (\ref{wyjsciowa_postac_ro_Killinga}) by $n^{\dot{B}}n^{\dot{D}}$ we obtain the condition
\begin{equation}
\label{rownanie_struny_ekspandujacej_anysam}
n^{\dot{B}} n^{\dot{D}} \nabla_{A \dot{B}} n_{\dot{D}} =0
\end{equation}
which means that the spinor $n_{\dot{B}}$ defines a congruence of anti-self-dual null strings given by the congruence of integral manifolds of the distribution $\{ r_{A}n_{\dot{B}}, s_{A} n_{\dot{B}} \}$, $r_{A}s^{A} \ne 0$. Substituting into the first equality of (\ref{definicja_eli}) the form of Killing vector $K_{A \dot{B}}=m_{A} n_{\dot{B}}$ then multiplying by $m^{A} m^{C}$ and using (\ref{rownanie_struny_ekspandujacej}) we get
\begin{equation}
\label{postac_lAB_dla_Killinga_zerowego}
m^{A} m^{C} \, l_{AC}=0 \ \Rightarrow \ l_{AC}=m_{(A}o_{C)}
\end{equation}
where $o_{C}$ is a nonzero spinor. Finally, inserting (\ref{postac_lAB_dla_Killinga_zerowego}) into the integrability condition (\ref{integrability_c_L}) with $\chi=0$ and multiplaying both sides by $m^{S}m^{T}$ one has
\begin{equation}
\label{warunek23}
(m^{T}o_{T}) (m^{S} \nabla_{R \dot{A}} m_{S})=0
\end{equation}
But, since for $\Lambda \ne 0$ all congruences of self-dual null strings are expanding (see subsection \ref{etykietka_subsekcji_21}) and by (\ref{rownanie_struny_ekspandujacej}) the spinoir $m_{A}$ defines a self-dual null string congruence the following condition holds true
\begin{equation}
\label{rownanie_struny_nieekspandujacej}
m^{S} \nabla_{R \dot{A}} m_{S} \ne 0
\end{equation}
Hence, (\ref{warunek23}) with (\ref{rownanie_struny_nieekspandujacej}) yield
\begin{equation}
m^{T}o_{T}=0 \ \Rightarrow \ o_{T}=\mu \, m_{T} \ \ \ , \ \ \ \mu \ne 0
\end{equation}
Therefore $l_{AB}=\mu \, m_{A}m_{B}$ and the proof of implication $(ii) \Rightarrow (iii)$ is completed. It remains to prove that $(iii) \Rightarrow (ii)$. Inserting $l_{ST}= \mu \, m_{S}m_{T}$ into the integrability condition (\ref{integrability_c_L}) and multiplaying by $m^{S}m^{T}$ we get
\begin{equation}
\frac{2}{3} \Lambda m_{R} m^{S} \, K_{S \dot{A}}=0 \ \stackrel{\Lambda \ne 0}{\Rightarrow} \ m^{S}K_{S \dot{A}} =0
\end{equation}
But this last equation implies the existence of some spinor $n_{\dot{A}}$ such that $K_{A\dot{B}}=m_{A}n_{\dot{B}}$. This ends the proof of implication $(iii) \Rightarrow (ii)$. So the proof of our lemma is complete. $\blacksquare$
\newline
\newline
Note that from (\ref{rownanie_struny_ekspandujacej_anysam}) and the complex counterpart of the Goldberg-Sachs theorem \cite{biblio_14, biblio_19} it follows that $C_{\dot{A}\dot{B}\dot{C}\dot{D}}$ is algebraically special and $n_{\dot{A}}$ is the multiple Penrose dotted spinor i.e.
\begin{equation}
C_{\dot{A}\dot{B}\dot{C}\dot{D}} \, n^{\dot{A}}n^{\dot{B}}n^{\dot{C}}=0
\end{equation}
Thus we have
\newline
\newline
\textbf{Lemma 4.3}
\newline
If $K_{A\dot{B}}=m_{A}n_{\dot{B}}$ is a null Killing vector then $C_{\dot{A}\dot{B}\dot{C}\dot{D}}$ is algebraically degenerate with $n_{\dot{A}}$ being a multiple Penrose dotted spinor and the space is of the type $[-] \otimes [\textrm{deg}]$. $\blacksquare$
\newline
\newline
From the lemma 4.2 one quickly finds
\newline
\newline
\textbf{Corollary 4.1}
\newline
If $K_{A\dot{B}}$ is a null Killing vector then at each point it is tangent to a self-dual null string and an anti-self-dual null string. These strings are defined by the spinor $m_{A}$ and $n_{\dot{B}}$, respectively, where $K_{A \dot{B}}=m_{A}n_{\dot{B}}$. $\blacksquare$
\newline
\newline
and
\newline
\newline
\textbf{Corollary 4.2}
\newline
$K_{A\dot{B}}$ is a nonnull Killing vector iff $l_{AB}l^{AB}=2 \det (l_{AB}) \ne 0$.
$\blacksquare$
\newline
\newline
From the proof of lemma 4.2 we can see that if $l_{AB}=\mu \, m_{A}m_{B}$ then the spinor $m_{A}$ defines a congruence of self-dual null strings since $m_{A}$ satisfies (\ref{rownanie_struny_ekspandujacej}). The question is what happens when $l_{AB}=m_{(A}o_{B)}$, $m_{A}o^{A} \ne 0$. The answer is
\newline
\newline
\textbf{Lemma 4.4}
\newline
Let $l_{AB}=m_{(A}o_{B)}$, $m_{A}o^{A} \ne 0$. Then both $m_{A}$ and $o_{B}$ define congruences of self-dual null strings.
\newline
\newline
\textbf{Proof}
Inserting $l_{ST}=m_{(S}o_{T)}$, $m_{S}o^{T} \ne 0$, into (\ref{integrability_c_L}) and multiplying both sides by $m^{R}m^{S}m^{T}$ we get Eq. (\ref{rownanie_struny_ekspandujacej}) Analogously, multiplying by $o^{R}o^{S}o^{T}$ one obtains $o^{R}o^{S} \nabla_{R \dot{A}} o_{S}=0$. Hence, both self-dual distributions $\{ m_{A}t_{\dot{B}}, m_{A}u_{\dot{B}} \}$ and $\{ o_{A}t_{\dot{B}}, o_{A}u_{\dot{B}} \}$ are completely integrable (by the Frobenius theorem \cite{biblio_21, biblio_dod_6}). This ends the proof. $\blacksquare$
\newline
\newline
Now we are ready to accomplish the classification of all $\mathcal{H}$-spaces with $\Lambda$ admitting a Killing symmetry.

\subsection{$\mathcal{H}$-spaces with $\Lambda$ admitting a null Killing vector.}
\label{subsekcja_o_zerowych}

Assume that $K_{A \dot{B}}$ is a null Killing vector on a $\mathcal{H}$-space with $\Lambda$. Then from Lemma 4.2 it follows that $K_{A \dot{B}}$ is of the form
\begin{equation}
\label{po_raz_kolejy}
K_{A \dot{B}} = m_{A} n_{\dot{B}}
\end{equation}
and $l_{AB}$ reads
\begin{equation}
\label{postac_l_AB_dla_zerrowegoK}
l_{AB}=\mu \, m_{A} m_{B} \ \ \ \ , \ \ \ \mu \ne 0
\end{equation}
where, by (\ref{rownanie_struny_ekspandujacej}), the 2-dimensional holomorphic distribution $\{ m_{A}t_{\dot{B}} , m_{A}u_{\dot{B}} \}$, $t_{\dot{B}}u^{\dot{B}} \ne 0$, is integrable and its integral 2-dimensional complex surfaces constitute the congruence of self-dual null strings; analogously, by (\ref{rownanie_struny_ekspandujacej_anysam}), the integral 2-dimensional surfaces of the distribution $\{ r_{A}n_{\dot{B}}, m_{A}u_{\dot{B}} \}$, $r_{A} s^{A} \ne 0$ constitute the congruence of anti-self-dual null strings. Consider the self-dual congruence. It is defined by the completely integrable Pfaff system \cite{biblio_dod_6}
\begin{equation}
\label{Pfaff_system}
m_{A} \, g^{A \dot{B}}=0 \ \ \ \ \ \dot{B}=\dot{1}, \dot{2}
\end{equation}
or, equivalently, by the following orthogonal 2-form $\Sigma$
\begin{equation}
\label{2_forma_definiujaca_uklad}
\Sigma=m_{A} g^{A\dot{1}} \wedge m_{C}g^{C \dot{2}} = \frac{1}{2} \in_{\dot{B}\dot{D}} m_{A}m_{C} \, g^{A\dot{B}} \wedge g^{C \dot{D}}
\end{equation}
Now we choose the null tetrad $(e^{1},e^{2},e^{3},e^{4})$ and the coordinates $(q_{\dot{A}}, p^{\dot{B}})$ introduced in subsection 2.1 so that our present congruence given by the Pfaff system (\ref{Pfaff_system}) is defined also by
\begin{equation}
\label{druga_definicja_dystrybucji}
dq_{\dot{A}}=0 \ \ \ \ , \ \ \ \dot{A}=\dot{1},\dot{2}
\end{equation}
Consequently, one has
\begin{equation}
\label{Proporcjonalnosc_sigmy}
\Sigma \sim dq_{\dot{1}} \wedge dq_{\dot{2}} \stackrel{\textrm{by } (\ref{tetrada_spinorowa})}{\sim} g^{2\dot{1}} \wedge g^{2\dot{2}}
\end{equation}
It means that the undotted spinor basis is taken so that the spinor $m_{A}$ has the components
\begin{equation}
\label{baza_spinorowa_dla_kongruencji}
m_{A} = (0, m_{2}) \ \ \ \ , \ \ \ m_{2} \ne 0
\end{equation}
Inserting (\ref{baza_spinorowa_dla_kongruencji}) into (\ref{postac_l_AB_dla_zerrowegoK}) we get
\begin{equation}
\label{specjalny_wybor_bazy_sspi}
l_{11}=0=l_{12}
\end{equation}
Employing the first equality of (\ref{specjalny_wybor_bazy_sspi}) in (\ref{postac_l11}) one finds
\begin{equation}
\label{zerowanie_sie_JDelta}
J_{\dot{N}} \delta^{\dot{N}} = 0
\end{equation}
Then from (\ref{zerowanie_sie_JDelta}), the second equality of (\ref{specjalny_wybor_bazy_sspi}) and (\ref{postac_l12}) with $\Lambda \ne 0$ one obtains
\begin{equation}
\label{zerowanie_sie_Kdelta_dla_zerowego}
K_{\dot{N}} \delta^{\dot{N}} = 0
\end{equation}
Therefore
\begin{equation}
\label{zerowanie_delty}
\delta^{\dot{N}}=0 \stackrel{\textrm{by } (\ref{rozklad_delty_na_aib})}{\Longleftrightarrow} a=0=b
\end{equation}
Substituting (\ref{zerowanie_delty}) into (\ref{integrability_condition_1}) and (\ref{integrability_condition_3}) with $\Lambda \ne 0$ we have
\begin{equation}
\label{zerowanie_alphy}
\alpha = 0
\end{equation}
and 
\begin{equation}
\label{zerowanie_xi}
\xi=0
\end{equation}
The transformation formula for $\epsilon$ (\ref{transformacja_epsilon}) under (\ref{zerowanie_delty}) and (\ref{zerowanie_alphy}) takes a simple form $\epsilon'=(\lambda w'_{w})^{-1} \epsilon$. Hence, without any loss of generality one can put
\begin{equation}
\label{postac_epsilon_dla_zerowegoKili}
\epsilon = - \frac{1}{\tau}
\end{equation}
Finally, inserting (\ref{zerowanie_delty}), (\ref{zerowanie_alphy}) and (\ref{postac_epsilon_dla_zerowegoKili}) into (\ref{ostateczna_postac_wektora_Kili}) and using (\ref{connection_miedzy_KiJ}) and (\ref{definicja_eta}) we get the Killing vector
\begin{equation}
\label{Killing_dle_null}
K=\frac{\partial}{\partial \eta}
\end{equation}
Now, from (\ref{transformacja_beta}) one infers that the function $\sigma=\sigma(w,t)$ can be chosen so that 
\begin{equation}
Y=0
\end{equation}
and this choice, with (\ref{zerowanie_delty}), (\ref{zerowanie_alphy}) and (\ref{postac_epsilon_dla_zerowegoKili}) satisfied, leads to 
\begin{equation}
\label{zerowanie_wielomianu}
\mathcal{P}=0
\end{equation}
where the function $\mathcal{P}$ is defined by (\ref{definicja_wielomianu_P}). Under (\ref{zerowanie_delty}), (\ref{zerowanie_alphy}), (\ref{zerowanie_xi}), (\ref{Killing_dle_null}) and (\ref{zerowanie_wielomianu}) the master equation (\ref{expanding_master_equation}) is brought to the extremally simple form
\begin{equation}
\frac{\partial W}{\partial \eta}=0 \Rightarrow W=W(\phi,w,t)
\end{equation}
Then the heavenly equation with $\Lambda$ (\ref{rownanie_niebianskie_zLambda_postac_ostateczna}) reads 
\begin{equation}
\label{zredukowane_rownanie_nieb}
W_{t\phi}+\frac{\Lambda}{6 \tau} W_{\phi\phi}=0
\end{equation}
The general solution of Eq. (\ref{zredukowane_rownanie_nieb}) is
\begin{equation}
\label{rozwiazanie_zred_row}
W(\phi,w,t) = F\Big( \phi - \frac{\Lambda t}{6 \tau} , w\Big) + f(w,t)
\end{equation}
where $F$ and $f$ are arbitrary functions of their arguments. From the formula (\ref{transformacja_funkcji_kluczowej}) for the gauge transformation of the key function $W$ one quickly concludes that without any los of generality the function $f(w,t)$ in (\ref{rozwiazanie_zred_row}) can be put zero. Substituting (\ref{rozwiazanie_zred_row}) into (\ref{metrykaprzestrzeni_HH_z_ekspanja}) we get the \textsl{general metric of a $\mathcal{H}$-space with $\Lambda$ admitting a null Killing vector}
\begin{eqnarray}
\label{main_metryka_null}
ds^2 &=& \phi^{-2} \bigg\{ \frac{2}{\tau} (d \eta \underset{s}{\otimes} d w - d \phi \underset{s}{\otimes} dt) + \frac{\Lambda}{3 \tau^{2}} \,  dt \underset{s}{\otimes}dt \ \ \ \ \ \ 
\\ \nonumber
&& \ \ \ \ \ \ \ \ \ \ 
  + 2  \left( 2  F_{\phi} - \phi \, F_{\phi \phi}   \right) dw \underset{s}{\otimes}dw \bigg\} \ \ \ \ \ \ 
\\ \nonumber
&&F=F \Big( \phi - \frac{\Lambda t}{6 \tau}, w \Big)
\end{eqnarray}
(Remark: $\tau \ne 0$ is any complex parameter and can be chosen as it is convenient). Substituting (\ref{rozwiazanie_zred_row}) into the second formula of (\ref{expanding_krzywizna}) one has
\begin{equation}
C_{\dot{A}\dot{B}\dot{C}\dot{D}}= \phi^{3} \, \frac{\partial^{4} F}{\partial \phi^{4}} J_{\dot{A}}J_{\dot{B}} J_{\dot{C}} J_{\dot{D}}
\end{equation}
Hence, if $F_{\phi\phi\phi\phi} \ne 0$ then $C_{\dot{A}\dot{B}\dot{C}\dot{D}}$ is of the type N and $J_{\dot{A}}$ is the multiple dotted Penrose spinor. Consequently, from Lemma 4.3 one infers that the spinor $n_{\dot{A}}$ defined in (\ref{po_raz_kolejy}) is proportional to $J_{\dot{A}}$
\begin{equation}
n_{\dot{A}} \sim J_{\dot{A}}
\end{equation}
So the null Killing vector $K=K^{a} \partial_{a}$ is a 4-fold Debever-Penrose vector
\begin{equation}
C_{abcd}K^{d}=0
\end{equation}
If $F_{\phi\phi\phi\phi}=0$ then $C_{\dot{A}\dot{B}\dot{C}\dot{D}}=0$ and the metric (\ref{main_metryka_null}) is a conformally flat complex Einstein metric with non-zero curvature scalar.

It is an easy matter to carry over all results of this subsection to the case of a real $\mathcal{H}$-space of signature $(++--)$ with $\Lambda$. Here the null Killing vector $K_{A \dot{B}}$ is real and one can quickly show that the spinors $m_{A}$ and $n_{\dot{B}}$ defined by (\ref{po_raz_kolejy}) can be chosen to be real spinors. Then the Lemmas 4.1, 4.2 and 4.3, and the results of subsection \ref{subsekcja_o_zerowych} hold true for a real $\mathcal{H}$-space of signature $(++--)$ with $\Lambda$ under the assumption that all complex objects (e.g. spinors, null strings, tetrads, coordinates, etc.) considered in the complex $\mathcal{H}$-space with $\Lambda$ are now real and instead of \textsl{holomorphic} objects we deal with \textsl{real} smooth objects.

In particular, the general metric of a real $\mathcal{H}$-space of signature $(++--)$ with $\Lambda$ is given by (\ref{main_metryka_null}) where $(\phi,\eta,w,t)$ are real coordinates, $\tau$ is a real parameter and $
F=F ( \phi - \frac{\Lambda t}{6 \tau}, w )$ is an arbitrary real smooth function. This metric is of some interest from the point of view of the Osserman geometry \cite{biblio_10}. Namely, it is the \textsl{general metric of signature $(++--)$ of the 4-dimensional globally Osserman space with non-zero curvature scalar admitting a null Killing vector}.

[Remark: We have learned from Maciej Dunajski that the same metric (\ref{main_metryka_null}) was recently found by himself and Paul Tod \cite{biblio_20}].

\subsection{$\mathcal{H}$-spaces with $\Lambda$ admitting a non-null Killing vector.}

Here we assume that the Killing vector $K_{A \dot{B}}$ on a $\mathcal{H}$-space with $\Lambda$ is non-null. From Corollary 4.2 we know, that this is equivalent to the statement that
\begin{equation}
\label{warunek_na_niezerowosc_Killinga}
l_{AB} l^{AB} = 2 \det (l_{AB}) \ne 0
\end{equation}
The condition (\ref{warunek_na_niezerowosc_Killinga}) is fullfilled iff
\begin{equation}
\label{l_AB_dla_niezerowego_Killinga}
l_{AB} = m_{(A}o_{B)} \ \ \ \ , \ \ \ \ m_{A}o^{A} \ne 0
\end{equation}
From Lemma 4.4 one concludes that 2-dimensional holomorphic distributions \linebreak $\{ m_{A} t_{\dot{B}}, m_{A} u_{\dot{B}} \}$ and $\{ o_{A} t_{\dot{B}}, o_{A} u_{\dot{B}} \}$, $t_{\dot{B}} u^{\dot{B}} \ne 0$ are completely integrable and their integral manifolds constitute two congruences of self-dual null strings. We take one of these congruences, for example that define by the Pfaff system (\ref{Pfaff_system}) or the 2-form (\ref{2_forma_definiujaca_uklad}). Then we choose the null tetrad $(e^{1}, e^{2}, e^{3}, e^{4} )$ and the coordinates $(p^{\dot{A}}, q_{\dot{B}})$ described in subsection 2.1 so that (\ref{druga_definicja_dystrybucji}), (\ref{Proporcjonalnosc_sigmy}) and (\ref{baza_spinorowa_dla_kongruencji}) are satisfied. Substituting (\ref{baza_spinorowa_dla_kongruencji}) into (\ref{l_AB_dla_niezerowego_Killinga}) one gets
\begin{equation}
\label{war_na_zerowaniesie_l11}
l_{11} = 0
\end{equation}
Hence, by (\ref{postac_l11}) we have (\ref{zerowanie_sie_JDelta}). From (\ref{warunek_na_niezerowosc_Killinga}) under (\ref{war_na_zerowaniesie_l11}) it follows that
\begin{equation}
\label{war_na_niezerowaniesie_l12}
l_{12} \ne 0
\end{equation}
Inserting (\ref{zerowanie_sie_JDelta}) into (\ref{postac_l12}) and using (\ref{war_na_niezerowaniesie_l12}) one concludes that (since $\Lambda \ne 0$)
\begin{equation}
\label{warunek_na_niezerowaniesie_Kdelta}
K_{\dot{N}} \delta^{\dot{N}} \ne 0 \stackrel{\textrm{by } (\ref{rozklad_delty_na_aib})}{\Longleftrightarrow} b \ne 0
\end{equation}
(compare with Eq. (\ref{zerowanie_sie_Kdelta_dla_zerowego}) which is fulfilled in the case of null Killing vector). Then from (\ref{integrability_condition_1}) and (\ref{integrability_condition_3}) under (\ref{zerowanie_sie_JDelta}) it follows that (\ref{zerowanie_alphy}) and (\ref{zerowanie_xi}) hold true.

The transformation rule (\ref{transformacja_b}) shows that since by (\ref{warunek_na_niezerowaniesie_Kdelta}) $b \ne 0$ one can choose the function $\lambda$ so that 
\begin{equation}
b'=1
\end{equation}
Finally, from (\ref{transformacja_epsilon}) and (\ref{transformacja_beta}) with $b \ne 0$ we conclude that the functions $\sigma$ and $M$ can be chosen so that 
\begin{equation}
\epsilon'=0 \ \ \ \ \textrm{and} \ \ \ \ Y'=0
\end{equation}
and one quickly gets from (\ref{definicja_wielomianu_P}) that
\begin{equation}
\mathcal{P}'=0
\end{equation}
Gathering all that we have that if the Killing vector $K_{A \dot{B}}$ is non-null then there exist the coordinates $(p^{\dot{A}}, q_{\dot{B}})$ or $(\phi, \eta,w,t)$ introduced in section 2 such that 
\begin{equation}
\label{zbior_uproszczen_dla_niezerowego_Kili}
a=\alpha=\xi=\epsilon=Y=\mathcal{P}=0 \ \ \ \ , \ \ \ \ b=1
\end{equation}
Inserting (\ref{zbior_uproszczen_dla_niezerowego_Kili}) into (\ref{ostateczna_postac_wektora_Kili}) one gets
\begin{equation}
\label{nonnnnulll_Kil}
K=\frac{\partial}{\partial t}
\end{equation}
Then the master equation (\ref{expanding_master_equation}) gives
\begin{equation}
\label{funkcja_kluczowa_dla_niezerowego_Kili}
\frac{\partial W}{\partial t}=0 \ \ \rightarrow \ \ W=W(\phi, \eta, w)
\end{equation}
Substituting (\ref{funkcja_kluczowa_dla_niezerowego_Kili}) into the heavenly equation with $\Lambda$ (\ref{rownanie_niebianskie_zLambda_postac_ostateczna}) we obtain
\begin{equation}
\label{zredukowane_HH_dla_niezerr}
 W_{\eta \eta}W_{\phi \phi} - W_{\eta \phi}^{2} + 2 \phi^{-1} \Big(  W_{\eta} W_{\eta \phi} -  W_{\phi}W_{\eta\eta}   \Big)
+ (\tau \phi)^{-1} \Big(  W_{w\eta}   - \frac{\Lambda}{6 \tau}  W_{\phi \phi}  \Big)  = 0 \ \ \ \
\end{equation}
We intend to bring Eq. (\ref{zredukowane_HH_dla_niezerr}) to more simple forms. To this end we first rewrite (\ref{zredukowane_HH_dla_niezerr}) in terms of differential forms
\begin{eqnarray}
\label{formowa_forma_rownania}
&&\phi \, dW_{\phi} \wedge dW_{\eta} \wedge dw + 2 W_{\eta} \, dW_{\eta} \wedge d\eta \wedge dw - 2W_{\phi} \, d\phi \wedge d W_{\eta} \wedge dw 
\\
\nonumber
&&+ \frac{1}{\tau} \, d\phi \wedge d\eta \wedge d W_{\eta} - \frac{\Lambda}{6 \tau^{2}} \, dW_{\phi} \wedge d \eta \wedge dw =0
\end{eqnarray}
Then one has
\begin{equation}
\label{rozzniczki}
dW = W_{\phi} \, d\phi + W_{\eta} \, d\eta +W_{w} \, dw \ \ \Rightarrow \ \ d (W-\eta W_{\eta}) = W_{\phi} \, d\phi - \eta \, dW_{\eta} + W_{w} \, dw
\end{equation}
In the next step we perform the Legendre transformation
\begin{eqnarray}
\label{transformata_Legendrea}
&&(\phi, \eta, w) \rightarrow (\phi, z, w) \ \ , \ \ z:= W_{\eta} \ \ \Rightarrow \ \ \eta = \eta(\phi, z, w) 
\\
\nonumber
&&V=V(\phi,z,w) := \frac{\Lambda}{6 \tau^{2}} \bigg[ W \Big( \phi, \eta (\phi, z, w), w \Big) - z \, \eta (\phi, z, w) \bigg]
\end{eqnarray}
From (\ref{rozzniczki}) and (\ref{transformata_Legendrea}) one quickly gets
\begin{equation}
\label{pocchodne}
W_{\phi} = \frac{6 \tau^{2}}{\Lambda} \, V_{\phi} \ \ , \ \ W_{w} = \frac{6 \tau^{2}}{\Lambda} \, V_{w} \ \ , \ \ \eta = - \frac{6 \tau^{2}}{\Lambda} \, V_{z}
\end{equation}
Inserting (\ref{transformata_Legendrea}) and (\ref{pocchodne}) into (\ref{formowa_forma_rownania}) and introducing
\begin{equation}
v := \tau w \ \ , \ \ U=U(\phi, z, v) := V(\phi, z, \tau^{-1} v)
\end{equation}
we obtain the following equation
\begin{eqnarray}
&&\phi \, dU_{\phi} \wedge dz \wedge dv - 2z \, dz \wedge dU_{z} \wedge dv - 2 U_{\phi} \, d\phi \wedge dz \wedge dv 
\\
\nonumber
&&- d \phi \wedge dU_{z} \wedge dz + d U_{\phi} \wedge d U_{z} \wedge dv = 0
\end{eqnarray}
which is equivalent to the nonlinear partial differential equation for $U$
\begin{equation}
\label{kolejna_redukcja}
U_{\phi \phi} U_{zz} - U_{z \phi}^{2} + \phi \, U_{\phi \phi} + 2z \, U_{z \phi} - 2 \, U_{\phi} + U_{zv}=0
\end{equation}
[Remark. Before we proceed further an important remark is needed. The Legendre transformation (\ref{transformata_Legendrea}) makes sense if $W_{\eta \eta} \ne 0$. So the case
\begin{equation}
\label{wetaetazero}
W_{\eta \eta}=0
\end{equation}
must be considered separately. Assume that Eq. (\ref{wetaetazero}) is fulfilled. The general solution of this equation $W=W(\phi, \eta, w)$ is of the form
\begin{equation}
\label{rozwiazanie_wetaetazero}
W= \eta \, f(\phi,w) + g(\phi,w)
\end{equation}
where $f=f(\phi, w)$ and $g=g(\phi,w)$ are arbitrary functions. Inserting (\ref{rozwiazanie_wetaetazero}) into the heavenly equation with $\Lambda$ (\ref{zredukowane_HH_dla_niezerr}) and performing straightforward manipulations one gets the general key function $W=W(\phi, \eta, w)$ satisfying (\ref{wetaetazero}) and (\ref{zredukowane_HH_dla_niezerr})
\begin{equation}
\label{kluczowa_dla_deSittera}
W=(f_{1}\phi +f_{2}) \, \eta + (f_{1}^{2} + \tau^{-1} f_{1}') \, \frac{\tau^{2}}{\Lambda} \phi^{3} + (2f_{1}f_{2} + \tau^{-1} f_{2}') \, \frac{3 \tau^{2}}{\Lambda} \phi^{2} +f_{3} \, \phi + f_{4}
\end{equation}
where $f_{1}$, $f_{2}$, $f_{4}$ and $f_{4}$ are arbitrary functions of the variable $w$ and $f_{1}':= \frac{d f_{1}}{dw}$, $f_{2}' := \frac{df_{2}}{dw}$. Substituting (\ref{kluczowa_dla_deSittera}) into the second formula of (\ref{expanding_krzywizna}) one easily concludes that now $C_{\dot{A}\dot{B}\dot{C}\dot{D}}=0$. Consequently, the case when (\ref{wetaetazero}) is fulfilled leads to the key function given by (\ref{kluczowa_dla_deSittera}) and the respective complex spacetime is conformally flat i.e. complex de Sitter].
\newline
Return to Eq. (\ref{kolejna_redukcja}). Simple substitutions
\begin{equation}
2P := U - \frac{1}{2} \phi z^{2} \ \ , \ \ y:=2v
\end{equation}
lead to the following equation for $P=P(\phi,z,y)$
\begin{equation}
\label{Kolejna_druga_redukcja}
P_{\phi\phi}P_{zz} - P_{z\phi}^{2} + \phi \, P_{\phi\phi} - P_{\phi} + P_{zy}=0
\end{equation}
This equation has exactly the same form as Eq. (4.11) of Ref. \cite{biblio_8}. Therefore, our further analysis of Eq. (\ref{Kolejna_druga_redukcja}) goes along the line described in \cite{biblio_8}. First, observe that (\ref{Kolejna_druga_redukcja}) can be equivalently presented in terms of 3-forms as
\begin{subequations}
\begin{eqnarray}
\label{Kolejna_trzecia_redukcja}
&&dr \wedge ds \wedge dy + \phi \, dr \wedge dz \wedge dy - r \, d\phi \wedge dz \wedge dy + d\phi \wedge dz \wedge ds
\\ 
\label{warunek_na_pochodne}
&& dr \wedge d \phi \wedge dy + ds \wedge dz \wedge dy =0 
\\ 
\label{warrrrrrunek}
&&d\phi \wedge dz \wedge dy \ne 0
\end{eqnarray}
\end{subequations}
[Indeed, from (\ref{warunek_na_pochodne}) it follows that there exists a function $P=P(\phi , z, y)$ such that $r=P_{\phi}$ and $s=P_{z}$. Then (\ref{Kolejna_trzecia_redukcja}) gives (\ref{Kolejna_druga_redukcja})].
\newline
Define the 1-form $\omega$ by
\begin{equation}
\label{definicja_formy_omega}
\omega := ds - r \, dy + \phi \, dz
\end{equation}
It is an easy matter to show that with the use of $\omega$ Eq. (\ref{Kolejna_trzecia_redukcja}) takes the form
\begin{equation}
\omega \wedge d \omega =0
\end{equation}
Consequently, by the Frobenius theorem \cite{biblio_21} one conlcudes that there exist functions $H=H(\phi,z,y)$ and $x=x(\phi,z,y)$ such that
\begin{equation}
\label{rozwiazanie_na_omega}
\omega = H  dx
\end{equation}
The exterior differentiation of the 1-form $\omega$ defined by (\ref{definicja_formy_omega}) gives $d\omega = -dr \wedge dy + d\phi \wedge dz$. Taking the exterior product $d \omega \wedge dy$ and using also (\ref{rozwiazanie_na_omega}) we get
\begin{equation}
\label{rownanie_prawie_juz_zred}
dH \wedge dx \wedge dy - d\phi \wedge dz \wedge dy =0
\end{equation}
Summing up the exterior products $d\omega \wedge d\phi$ and $\omega \wedge dz \wedge dy$, and using Eq. (\ref{warunek_na_pochodne}) one obtains
\begin{equation}
\label{kolejnyformowywarunek}
d \omega \wedge d\phi + \omega \wedge dz \wedge dy =0 \ \ \stackrel{\textrm{by } (\ref{rozwiazanie_na_omega})}{\Longrightarrow} \ \ dH \wedge dx \wedge d \phi + H \, dx \wedge dz \wedge dy =0
\end{equation}
First, we can show that
\begin{equation}
\label{warunek_na_zmienne_niezalezne}
dx \wedge d\phi \wedge dy \ne 0
\end{equation}
Indeed, from (\ref{definicja_formy_omega}) with (\ref{rozwiazanie_na_omega}) it follows that $H \, dx \wedge d \phi \wedge dy=0$ iff $ds \wedge d\phi \wedge dy + \phi \, dz \wedge d\phi \wedge dy=0$. Then substituting $s=P_{z}$ and remembering that $d\phi \wedge dz \wedge dy \ne 0$ one concludes that $H \, dx \wedge d\phi \wedge dy=0$ iff $P_{zz} = -\phi$, i.e. $P=-\frac{1}{2} \phi z^{2} + z A(\phi, y) + B(\phi, y)$ where $A=A(\phi, y)$ and $B=B(\phi, y)$ are some functions of $(\phi, y)$. However, such a $P$ does not satisfy Eq. (\ref{Kolejna_druga_redukcja}). Therefore, the condition (\ref{warunek_na_zmienne_niezalezne}) holds true. Consequently, $(x, \phi, y)$ are independent variables. Eq. (\ref{rownanie_prawie_juz_zred}) is then equivalent to the statement that
\begin{equation}
\label{Red1}
z_{x} = H_{\phi}
\end{equation}
and Eq. (\ref{kolejnyformowywarunek}) is equivalent to
\begin{equation}
\label{Red2}
-z_{\phi} = (\ln H)_{y}
\end{equation}
From (\ref{Red1}) and (\ref{Red2}), substituting also
\begin{equation}
\label{definijcaF_jakologarytn_zH}
F := \ln H
\end{equation}
one arrives at the \textsl{Boyer - Finley - Plebański (or the Toda field) equation} \cite{biblio_8}
\begin{equation}
\label{rownanie_BFP}
F_{xy} + (e^{F})_{\phi \phi} = 0
\end{equation}
From (\ref{rownanie_prawie_juz_zred}), (\ref{warunek_na_zmienne_niezalezne}) and (\ref{definijcaF_jakologarytn_zH}), under (\ref{warrrrrrunek}) one infers that
\begin{equation}
F_{\phi} \ne 0
\end{equation}

\subsection{Comments on real $\mathcal{H}$-spaces with $\Lambda$ admitting a non-null Killing vector.}

Real $\mathcal{H}$-spaces of Euclidian signature $(++++)$ with $\Lambda$ admitting a Killing vector have been studied in Refs. \cite{biblio_11, biblio_12}. In particular  in \cite{biblio_12} it has been proved that the problem can be always reduced to the solution of the Boyer - Finley - Plebański equation. The case of a real $\mathcal{H}$-space of signature $(++--)$ with $\Lambda$ admitting a non-null Killing vector is much more involved as now one must consider two different cases, namely
\begin{equation}
l_{AB}l^{AB} > 0 \ \ \ \ \textrm{or} \ \ \ \ l_{AB}l^{AB} < 0 
\end{equation}
This problem has been solved by M. H\"ogner \cite{biblio_13} and it has been shown that there exist suitable coordinates such that again one arrives at the Boyer - Finley - Plebański equation. We do not enter these questions here, but we note only that the real case with $l_{AB}l^{AB} <0$ can be easily obtained from subsection 4.3 by taking real objects instead of the holomorphic ones.

\appendix

\renewcommand{\thesection}{\Alph{section}}
\renewcommand{\theequation}{\Alph{section}.\arabic{equation}}

\setcounter{equation}{0}
\setcounter{section}{0}

\setlength\arraycolsep{2pt}
\setcounter{equation}{0}

\section{Appendix.}

\subsection{Explicit form of the Killing equations and their integrability conditions in $\mathcal{H}$-spaces with $\Lambda$.}
\label{sekcja_jawna_postac_rownan}

Equations (\ref{rozklad_rownania_Killinga}) - (\ref{definicja_chi}) together with their integrability conditions (\ref{integrability_c_L}) form our problem to be solved. Using the formula for the spinorial covariant derivative
\begin{eqnarray}
\nabla_{M \dot{N}} \Psi^{A \dot{B}}_{C \dot{D}} &=& \partial_{M \dot{N}} \Psi^{A \dot{B}}_{C \dot{D}} 
+ \mathbf{\Gamma}^{A}_{\ SM \dot{N}} \, \Psi^{S \dot{B}}_{C \dot{D}} 
- \mathbf{\Gamma}^{S}_{\ CM \dot{N}} \, \Psi^{A \dot{B}}_{S \dot{D}}
\\ \nonumber
&&+ \mathbf{\Gamma}^{\dot{B}}_{\ \dot{S}M \dot{N}} \, \Psi^{A \dot{S}}_{C \dot{D}}
- \mathbf{\Gamma}^{\dot{S}}_{\ \dot{D}M \dot{N}} \, \Psi^{A \dot{B}}_{C \dot{S}}
\end{eqnarray}
then the decomposition (\ref{rozklad_koneksji}), the formulae (\ref{rozklad_operatora}) and (\ref{rozklad_wektora_Killinga}), after some work, one obtains the equations
\begin{eqnarray}
\label{rowwnania_Killinga}
E_{11}^{\ \ \dot{A}\dot{B}} &\equiv& 2 \phi^{-2} \, \partial^{(\dot{A}} \big( \phi^{2} k^{\dot{B})} \big) = 0
\\ \nonumber
E_{12}^{\ \ \dot{A}\dot{B}} &\equiv& \partial^{(\dot{A}} \big( h^{\dot{B})} - \phi^{2}k_{\dot{S}} \, Q^{\dot{B})\dot{S}} \big) +\phi^{2} \, \frac{\partial k^{(\dot{A}}}{\partial q_{\dot{B})}} + Q^{\dot{A}\dot{B}} \, \partial^{\dot{S}} (\phi^{2}k_{\dot{S}})=0
\\ \nonumber
E_{22}^{\ \ \dot{A}\dot{B}} &\equiv& 2 \, \eth^{(\dot{A}}h^{\dot{B})} + 2\phi^{2} \, \eth_{\dot{S}}Q^{\dot{S}(\dot{A}} k^{\dot{B})} - 2\phi^{2} \, h^{\dot{S}} \partial_{\dot{S}} Q^{\dot{A}\dot{B}}=0
\\ \nonumber
\frac{1}{2} E &\equiv&  -\eth^{\dot{N}} k_{\dot{N}} + \partial^{\dot{N}} h_{\dot{N}} + 4\phi^{-1} h_{\dot{S}}J^{\dot{S}} - \phi^{4} k_{\dot{S}} \, \partial_{\dot{A}} (\phi^{-2} Q^{\dot{A}\dot{S}})=0
\ \ \ \ \ \ \ \ \ \ \ \ \ \ \ \ \ \ \ \ \ \ \ \ \  \ \ \ \ \ \ 
\end{eqnarray}
then
\begin{eqnarray}
\label{definicje_spinorow_LAB}
l_{11} &=& \partial^{\dot{N}} k_{\dot{N}}
\\ \nonumber
2 \, l_{12} &=& \partial_{\dot{N}} h^{\dot{N}} + \eth_{\dot{N}} k^{\dot{N}} +2\phi^{-1} \, h_{\dot{N}}J^{\dot{N}} +\phi^{2} k_{\dot{N}} \, \partial_{\dot{S}} Q^{\dot{S}\dot{N}}
\\ \nonumber
l_{22} &=& \eth_{\dot{N}} h^{\dot{N}} +\phi^{2} k_{\dot{N}} \, \eth_{\dot{S}} Q^{\dot{S}\dot{N}} - 2\phi\, h_{\dot{N}} J_{\dot{M}} \, Q^{\dot{N}\dot{M}}
\ \ \ \ \ \ \ \ \ \ \ \ \ \ \ \ \ \ \ \ \ \ \ \ \ \ \ \ \ \ \ \ \ \ \ \ \ \ \ \ \ \ \ \ \ \ \ \ \ \ 
\end{eqnarray}
and the integrability conditions $L_{RST}^{\ \ \ \ \; \dot{A}}$
\begin{eqnarray}
-\frac{1}{\sqrt{2}} \, L_{111}^{\ \ \ \; \dot{A}} &\equiv&  \phi^{-3} \, \partial^{\dot{A}} \big( \phi^{3} l_{11} \big) = 0
\\ \nonumber
-\frac{1}{\sqrt{2}} \, L_{112}^{\ \ \ \; \dot{A}} &\equiv&  \partial^{\dot{A}} l_{12} + \phi \, l_{11} \, J_{\dot{S}}Q^{\dot{S}\dot{A}} + \frac{\Lambda}{3} \, k^{\dot{A}} = 0
\\ \nonumber
-\frac{1}{\sqrt{2}} \, L_{122}^{\ \ \ \; \dot{A}} &\equiv&  \partial^{\dot{A}} l_{22} +3\phi^{-1} \, l_{22} J^{\dot{A}} + 2\phi \, l_{12} J_{\dot{S}}Q^{\dot{S}\dot{A}}  +    \frac{2}{3} \Lambda \, h^{\dot{A}}    =0
\\ \nonumber
-\frac{1}{\sqrt{2}} \, L_{211}^{\ \ \ \; \dot{A}} &\equiv&  \eth^{\dot{A}} l_{11} + \phi \, l_{11} \, \partial_{\dot{S}}(\phi Q^{\dot{S}\dot{A}} ) - 2\phi^{-1} \, l_{12} \, J^{\dot{A}} -   \frac{2}{3} \Lambda \, k^{\dot{A}}   = 0
\\ \nonumber
-\frac{1}{\sqrt{2}} \, L_{212}^{\ \ \ \; \dot{A}} &\equiv&  \eth^{\dot{A}} l_{12} - \phi^{-1} \, l_{22} \, J^{\dot{A}} +\phi^{2} \, l_{11} \, \eth_{\dot{S}}Q^{\dot{S}\dot{A}} -   \frac{\Lambda}{3} \, h^{\dot{A}}  =0
\\ \nonumber
-\frac{1}{\sqrt{2}} \, L_{222}^{\ \ \ \; \dot{A}} &\equiv&  \eth^{\dot{A}} l_{22} -\phi \, l_{22} \, \partial_{\dot{S}}(\phi Q^{\dot{S}\dot{A}} ) + 2 \phi^{2} \, l_{12} \, \eth_{\dot{S}} Q^{\dot{S}\dot{A}} =0
\end{eqnarray}

\subsection{Detailed derivation of the master equation.}
\label{sekcja_szczegolowe_obliczenia}

In this section we present in some details the reduction of Killing equations to one master equation (\ref{expanding_master_equation}).  From the Killing equations $E_{11}^{\ \ \dot{A}\dot{B}}$ (see (\ref{rowwnania_Killinga})), one finds that 
\begin{equation}
k^{\dot{A}} = \phi^{-2} \, (\alpha p^{\dot{A}} + \delta^{\dot{A}})
\end{equation}
where $\alpha=\alpha(q^{\dot{M}})$ and $\delta^{\dot{A}} = \delta^{\dot{A}} (q^{\dot{M}})$ are arbitrary functions of their variables. Using this form of $k^{\dot{A}}$ one can find $l_{11}$ (\ref{postac_l11}) and check, that integrability condition $L_{111}^{\ \ \ \; \dot{A}}$ is automatically satisfied. Knowing the form of $k_{\dot{A}}$, the second triplet of Killing equations $E_{12}^{\ \ \dot{A}\dot{B}}$ can be brought to the form
\begin{equation}
E_{12}^{\ \ \dot{A}\dot{B}} \equiv \partial^{(\dot{A}} V^{\dot{B})}=0 
\end{equation}
where
\begin{equation}
V^{\dot{A}} := h^{\dot{A}} - \phi^{2} k_{\dot{S}} \, Q^{\dot{S}\dot{A}} + 2\alpha \, \Omega^{\dot{A}} + \frac{\partial \alpha}{\partial q^{\dot{S}}} \, p^{\dot{S}}p^{\dot{A}} + \frac{\partial \delta^{\dot{S}}}{\partial q_{\dot{A}}} \, p_{\dot{S}}
\end{equation}
But if $\partial^{(\dot{A}} V^{\dot{B})}=0$ then $V^{\dot{A}} = V p^{\dot{A}} + \epsilon^{\dot{A}}$ with $V=V(q^{\dot{M}})$ and $\epsilon^{\dot{A}} = \epsilon^{\dot{A}} (q^{\dot{M}})$ being arbitrary functions of their variables. Hence
\begin{equation}
\label{postac_h}
h^{\dot{A}} - \phi^{2} k_{\dot{S}} \, Q^{\dot{S}\dot{A}} =
- 2\alpha \, \Omega^{\dot{A}} - \frac{\partial \alpha}{\partial q^{\dot{S}}} \, p^{\dot{S}}p^{\dot{A}}
 - \frac{\partial \delta^{\dot{S}}}{\partial q_{\dot{A}}} \, p_{\dot{S}}
 + V  p^{\dot{A}} + \epsilon^{\dot{A}}
\end{equation}
The final Killing equation, i.e., the $E$ equation can be rearranged to the form
\begin{equation}
\label{E_equation_in_different_form}
\frac{1}{2} E \equiv \phi^{4} \, \partial_{\dot{N}} \Big[ \phi^{-4} \big( h^{\dot{N}} - \phi^{2} k_{\dot{S}} \, Q^{\dot{S}\dot{N}} \big) \Big]  - \frac{\partial }{\partial q^{\dot{N}}} (\alpha p^{\dot{N}} + \delta^{\dot{N}}) = 0
\end{equation}
Inserting $h^{\dot{A}} - \phi^{2} k_{\dot{S}} \, Q^{\dot{S}\dot{A}}$ from (\ref{postac_h}) into (\ref{E_equation_in_different_form}) and using the definition of $\Omega^{\dot{N}}$ one gets the polynomial in $p^{\dot{N}}$, and, finally, integrability condition (\ref{integrability_condition_1}) and the solutions for $V$ and $\epsilon^{\dot{A}}$
\begin{eqnarray}
&& V=  \frac{2}{\tau} \, K_{\dot{S}}J_{\dot{N}} \, \frac{\partial \delta^{\dot{S}}}{\partial q_{\dot{N}}}
\\ \nonumber
&&J_{\dot{N}}\epsilon^{\dot{N}} = 0 \ \ \ \rightarrow \ \ \ \epsilon^{\dot{N}} = \epsilon \, J^{\dot{N}}
\end{eqnarray} 
with $\epsilon = \epsilon(q^{\dot{M}})$ being an arbitrary function. Putting this into (\ref{postac_h}) one obtains the final solution for $h^{\dot{A}}$
\begin{subequations}
\begin{eqnarray}
\label{solution_for_h}
&&h^{\dot{A}} - \phi^{2} k_{\dot{S}} \, Q^{\dot{S}\dot{A}} = -2\alpha \, \Omega^{\dot{A}} + R^{\dot{A}}
\\
\label{definicja_R_A}
&&R^{\dot{A}} := - \frac{\partial \alpha}{\partial q^{\dot{S}}} \, p^{\dot{S}}p^{\dot{A}}
 + \frac{\partial \delta_{\dot{S}}}{\partial q_{\dot{A}}} \, p^{\dot{S}}
 +  \frac{2}{\tau} \, K_{\dot{S}}J_{\dot{N}} \, \frac{\partial \delta^{\dot{S}}}{\partial q_{\dot{N}}} \,  p^{\dot{A}} + \epsilon J^{\dot{A}}
\end{eqnarray}
\end{subequations}
With $h^{\dot{A}}$ given by (\ref{solution_for_h}) one can calculate $l_{12}$ and obtain (\ref{postac_l12}). Integrability conditions $L_{112}^{\ \ \ \; \dot{A}}$ and $L_{211}^{\ \ \ \; \dot{A}}$ become identities.

The next step is to calculate the form of $l_{22}$ from $L_{212}^{\ \ \ \; \dot{A}}$. In $L_{212}^{\ \ \ \; \dot{A}}$ it is useful to replace the factor $\partial_{\dot{B}} l_{12}$ from $L_{112}^{\ \ \ \; \dot{A}}$ equation and do not calculate it directly from (\ref{postac_l12}). Except this, there appears the factor $Q^{\dot{A}\dot{B}}Q^{\dot{X}}_{\ \; \dot{B}}$ which is skew symmetric in the indices $\dot{A}\dot{X}$ and it can be changed according to
\begin{equation}
Q^{\dot{A}\dot{B}}Q^{\dot{X}}_{\ \; \dot{B}} = \frac{1}{2} \in^{\dot{A}\dot{X}}  Q^{\dot{S}\dot{B}}Q_{\dot{S}\dot{B}} =  \in^{\dot{A}\dot{X}} \phi^{2} \, \mathcal{T}
\end{equation}
Using the heavenly equation with $\Lambda$ in the form (\ref{expanding_hyperheavenly_equation_form1}) to replace $\mathcal{T}$ one can remove quadratic terms with the second derivative of the key function $W$. After some algebraic work, using (\ref{integrability_condition_1}) and contracting $L_{212}^{\ \ \ \; \dot{A}}$ with $K_{\dot{A}}$ one obtains $l_{22}$ in the form (\ref{postac_l22}) (contraction $J_{\dot{A}} \cdot L_{212}^{\ \ \ \; \dot{A}}$ is an identity).

The integrability condition $L_{122}^{\ \ \ \; \dot{A}}$ gives no additional information since it becomes an identity.

Eq. (\ref{postac_l22}) must be consistent with the definition of $l_{22}$ given by (\ref{definicje_spinorow_LAB}). It is not an identity. Denote
\begin{equation}
\label{Sigma}
\Sigma := -\pounds_{K} W + 4 \alpha \Upsilon - W \bigg( - 3 \frac{\partial \alpha}{\partial q^{\dot{S}}} \, p^{\dot{S}}  + \frac{1}{\tau}  \, \frac{\partial \delta^{\dot{S}}}{\partial q_{\dot{N}}} (3 J_{\dot{N}}K_{\dot{S}} + 2 K_{\dot{N}} J_{\dot{S}} ) \bigg)
\end{equation}
After some tedious calculations one shows that the consistency condition takes the form
\begin{eqnarray}
\label{consistency_condition_between_l22}
J_{\dot{A}} \, \partial^{\dot{A}} \Sigma &=&  \frac{1}{2} \frac{\partial^{2} \alpha}{\partial q^{\dot{N}} \partial q^{\dot{S}}} \, p^{\dot{N}} p^{\dot{S}} -\frac{1}{2} J^{\dot{N}} \, \frac{\partial \epsilon}{\partial q^{\dot{N}}}  - \frac{\Lambda}{6 \tau^{2}} \, \frac{\partial \delta^{\dot{S}}}{\partial q_{\dot{N}}} \, K_{\dot{N}}K_{\dot{S}}
\\ \nonumber
&&-\frac{1}{2 \tau} \, \frac{\partial^{2} \delta^{\dot{S}}}{\partial q_{\dot{N}} \partial q_{\dot{R}}} \, (2K_{\dot{S}}J_{\dot{N}} + J_{\dot{S}}K_{\dot{N}}) p_{\dot{R}} 
 + \frac{\Lambda}{6 \tau^{2}} \frac{\partial \alpha}{\partial q^{\dot{S}}} K^{\dot{S}} \eta 
\end{eqnarray}
Eq. (\ref{consistency_condition_between_l22}) plays a crucial role in the most important part of this work i.e. in an integration of the third triplet of the Killing equations $E_{22}^{\ \ \dot{A}\dot{B}}$. Putting $h^{\dot{A}}$ given by (\ref{solution_for_h}) into $E_{22}^{\ \ \dot{A}\dot{B}}$, after some obvious cancellations one gets
\begin{eqnarray}
\frac{1}{2} \phi^{-2} E_{22}^{\ \ \dot{A}\dot{B}} &\equiv& \frac{\partial R^{(\dot{A}}}{\partial q_{\dot{B})}} + Q^{\dot{S}(\dot{A}} \, \partial_{\dot{S}}R^{\dot{B})} - R^{\dot{S}} \, \partial_{\dot{S}} Q^{\dot{A}\dot{B}} + \frac{\partial (\phi^{2} k_{\dot{S}})}{\partial q_{(\dot{A}}} \, Q^{\dot{B})\dot{S}} + \phi^{2} k^{\dot{S}} \, \frac{\partial Q^{\dot{A}\dot{B}}}{\partial q^{\dot{S}}} \ \ \ \ \ \ \ \ \ 
\\ \nonumber
&&-2 \, \frac{\partial \alpha}{\partial q_{(\dot{A}}} \, \Omega^{\dot{B})} - 2 \alpha \, \frac{\partial \Omega^{(\dot{A}}}{\partial q_{\dot{B})}} - 2 \alpha \, Q^{\dot{S}(\dot{A}} \partial_{\dot{S}} \Omega^{\dot{B})} + 2 \alpha \, \Omega^{\dot{S}} \partial_{\dot{S}} Q^{\dot{A}\dot{B}} = 0
\end{eqnarray}
with $R^{\dot{A}}$ defined by (\ref{definicja_R_A}).Taking $Q^{\dot{A}\dot{B}}$ in the form (\ref{expanding_QAB_z_Omega}) we obtain
\begin{equation}
\frac{1}{2} \phi^{-2} E_{22}^{\ \ \dot{A}\dot{B}} \equiv \partial^{(\dot{A}} \Sigma^{\dot{B})} = 0
\end{equation}
where
\begin{eqnarray}
\label{Sigma_dotB}
\Sigma^{\dot{B}} &:=& -R^{\dot{S}} \, \partial_{\dot{S}} \Omega^{\dot{B}} + 2\Omega^{\dot{B}} \,  \frac{2}{\tau} K_{\dot{S}} J_{\dot{N}} \frac{\partial \delta^{\dot{S}}}{\partial q_{\dot{N}}} 
+\frac{\partial (\phi^{2} k^{\dot{N}})}{\partial q^{\dot{N}}} \, \Omega^{\dot{B}} 
+ \frac{\partial (\phi^{2}k_{\dot{S}})}{\partial q_{\dot{B}}} \, \Omega^{\dot{S}}
\\ \nonumber
&&  + \phi^{2} k^{\dot{S}} \, \frac{\partial \Omega^{\dot{B}}}{\partial q^{\dot{S}}}
+2\alpha \, \Omega^{\dot{S}} \, \partial_{\dot{S}} \Omega^{\dot{B}}+12\alpha \Upsilon J^{\dot{B}} + \frac{2}{\tau} \Lambda \alpha W K^{\dot{B}} - 6 \alpha \phi J^{\dot{S}} W_{p^{\dot{S}}}W_{p_{\dot{B}}}
\\ \nonumber
&&+3\alpha \phi \, \frac{\partial W}{\partial q_{\dot{B}}} -4 \frac{\partial \alpha}{\partial q^{\dot{S}}} \, p^{\dot{S}} \, \Omega^{\dot{B}} + J^{\dot{B}} \, \frac{\partial \epsilon}{\partial q^{\dot{S}}} \, p^{\dot{S}}
-\frac{1}{2} \frac{\partial^{2} \alpha}{\partial q^{\dot{S}} \partial q^{\dot{R}}} \, p^{\dot{S}} p^{\dot{R}} p^{\dot{B}}
\\ \nonumber
&& -\frac{\partial^{2} \delta^{\dot{S}}}{\partial q_{\dot{B}} \partial q_{\dot{R}}} \, p_{\dot{S}}p_{\dot{R}} + \frac{1}{2} \frac{\partial^{2} \delta^{\dot{B}}}{\partial q_{\dot{S}} \partial q_{\dot{R}}} \, p_{\dot{S}}p_{\dot{R}} + \frac{\partial}{\partial q^{\dot{T}}} \bigg(  \frac{2}{\tau} K_{\dot{S}} J_{\dot{N}} \frac{\partial \delta^{\dot{S}}}{\partial q_{\dot{N}}} \bigg)    p^{\dot{T}}p^{\dot{B}}
\end{eqnarray}
Inserting $\Omega^{\dot{B}}$ defined by (\ref{postac_Omegi}) into (\ref{Sigma_dotB}) one observes that the terms containing the key function $W$ in (\ref{Sigma_dotB}) can be rearranged into the form $\phi^{4} \partial^{\dot{B}}(\phi^{-3}\Sigma )$ with $\Sigma$ given by (\ref{Sigma}). 
Moreover, after some algebraic tricks one finds that
\begin{eqnarray}
&&-\frac{\partial^{2} \delta^{\dot{S}}}{\partial q_{\dot{B}} \partial q_{\dot{R}}} \, p_{\dot{S}}p_{\dot{R}} + \frac{1}{2} \frac{\partial^{2} \delta^{\dot{B}}}{\partial q_{\dot{S}} \partial q_{\dot{R}}} \, p_{\dot{S}}p_{\dot{R}} + \frac{\partial}{\partial q^{\dot{T}}} \bigg(  \frac{2}{\tau} K_{\dot{S}} J_{\dot{N}} \frac{\partial \delta^{\dot{S}}}{\partial q_{\dot{N}}} \bigg)    p^{\dot{T}}p^{\dot{B}}
=
\\ \nonumber
&&=\phi^{4} \, \partial^{\dot{B}} \bigg[ \phi^{-4} \frac{1}{2} \frac{\partial^{2} \delta^{\dot{S}}}{\partial q_{\dot{T}} \partial q_{\dot{R}}} \, p_{\dot{R}} p_{\dot{S}} p_{\dot{T}} \bigg]
-\frac{2}{\tau} \frac{\partial^{2} \delta^{\dot{S}}}{\partial q_{\dot{N}} \partial q_{\dot{T}}} J_{\dot{S}}J_{\dot{N}}p_{\dot{T}}p^{\dot{B}} \eta \phi^{-1}
\end{eqnarray}
Except this, from $\partial^{(\dot{A}} \Sigma^{\dot{B})} = 0$ it follows that $\Sigma^{\dot{B}} = X \, p^{\dot{B}} + Y^{\dot{B}}$ (with $X$ and $Y^{\dot{B}}$ being an arbitrary functions of the variable $q^{\dot{M}}$). Finally
\begin{eqnarray}
\label{Sigma_2}
&&X p^{\dot{B}} + Y^{\dot{B}} = 
\\ \nonumber
&&\phi^{4} \partial^{\dot{B}} \bigg( \phi^{-3}\Sigma   +\frac{1}{2} \phi^{-4}  \frac{\partial^{2} \delta^{\dot{S}}}{\partial q_{\dot{T}} \partial q_{\dot{R}}} \, p_{\dot{R}} p_{\dot{S}} p_{\dot{T}} 
\\ \nonumber
 && \ \ \ \ \ \ \ \ \
- \frac{1}{2 \tau^{3}} \frac{\partial^{2} \alpha}{\partial q^{\dot{S}} \partial q^{\dot{R}}} \eta \phi^{-3} \Big( \phi^{2}K^{\dot{S}}K^{\dot{R}} + \eta \phi K^{\dot{S}} J^{\dot{R}} + \frac{1}{3} \eta^{2} J^{\dot{R}}J^{\dot{S}} \Big) 
+ \frac{\Lambda}{6 \tau^{2}} \frac{\partial \alpha}{\partial q^{\dot{S}}} p^{\dot{S}} \eta^{2} \phi^{-4} \bigg) 
\\ \nonumber
&& + J^{\dot{B}} \, \frac{\partial \epsilon}{\partial q^{\dot{S}}} p^{\dot{S}} - \frac{\Lambda \epsilon}{6 \tau} K^{\dot{B}} + \frac{\Lambda}{6 \tau^{2}}  \frac{1}{\tau} \frac{\partial \delta^{\dot{S}}}{\partial q_{\dot{N}}} K_{\dot{S}} K_{\dot{N}} \big( \eta J^{\dot{B}} - \phi K^{\dot{B}} \big) 
\end{eqnarray}
Contracting (\ref{Sigma_2}) with $J_{\dot{B}}$ and using (\ref{consistency_condition_between_l22}) we get
\begin{eqnarray}
\label{solutions_for_omega_and_beta}
-X &=&  \frac{1}{2} J^{\dot{N}} \, \frac{\partial \epsilon}{\partial q^{\dot{N}}} + \frac{\Lambda}{3 \tau^{2}}
 \frac{\partial \delta^{\dot{S}}}{\partial q_{\dot{N}}} K_{\dot{S}}K_{\dot{N}}
\\ \nonumber
-Y^{\dot{B}} &=& \frac{\Lambda \epsilon}{6\tau} \, K^{\dot{B}} + \frac{1}{\tau} Y J^{\dot{B}}
\end{eqnarray}
where $Y=Y(q^{\dot{M}})$ is an arbitrary function.
\newline
(An alternative way to obtain (\ref{solutions_for_omega_and_beta}) is to multiply (\ref{Sigma_2}) by $\phi^{-4}$ and derive $\partial_{\dot{B}} (\ref{Sigma_2})$).
Using the solutions for $X$ and $Y^{\dot{B}}$ in Eq. (\ref{Sigma_2}) one can bring it to the form $\phi^{4} \partial^{\dot{B}} [\phi^{-3} (...)]=0$ and finally obtain the master equation.

Integrability conditions $L_{222}^{\ \ \ \; \dot{A}}$ give (\ref{integrability_condition_3}).

\end{document}